\documentclass[english,nofootinbib,twocolumn,rmp,aps]{revtex4-1}
\usepackage{graphicx}
\usepackage{dcolumn}
\usepackage{bm}
\usepackage{hyperref}
\renewcommand{\vec}[1]{{\mathbf{#1}}}
\renewcommand{\cite}{\citep}
\newcommand{\da}{\downarrow}
\newcommand{\ua}{\uparrow}
\newcommand{\beq}{\begin{eqnarray}}
\newcommand{\eeq}{\end{eqnarray}}
\newcommand{\bs}{\bar\sigma}

\newcommand{\tn}{\tilde n}
\begin{document}
\title{Mottness: Identifying the Propagating Charge Modes
in doped Mott Insulators}
\author{Philip Phillips}
\affiliation{Department of Physics,
University of Illinois
1110 W. Green Street, Urbana, IL 61801, U.S.A.}
\date{\today}

\begin{abstract}
High-temperature superconductivity in the copper-oxide ceramics
remains an unsolved problem because we do not know what the
propagating degrees of freedom are in the normal state.  As a result,
we do not know what are the weakly interacting degrees of freedom
which pair up to form the superconducting condensate. That the
electrons are not the propagating degrees of freedom in the cuprates
is seen most directly from experiments that show spectral weight
redistributions over all energy scales. Of course, the actual propagating degrees of freedom minimize such spectral rearrangements.  This review focuses on the
range of epxerimental consequences such UV-IR mixings have on the
normal state of the cuprates, such as the pseudogap,
mid-infrared band, temperature dependence of the Hall number, the
superfluid density,
and a recent theoretical advance which permits the identification of
the propagating degrees of freedom in a doped Mott insulator.  Within this theory, we
show how the wide range of phenomena which typify the normal state of
the cuprates arises including $T-$linear resistivity.
\end{abstract}

\maketitle
\tableofcontents

\section{Introduction}

The secret to solving any many-body problem is to correctly identify
the propagating degrees of freedom. Typically the propagating modes
cannot be read off by inspecting a Hamiltonian but rather are
dynamically generated through
a collective organization of the elemental fields.  In identifying the
principle that leads to such organization, it helps to know what to
throw out.  In this context,
the Bardeen-Cooper-Schrieffer (BCS)
theory\cite{bcs} of superconductivity in ordinary metals is remarkable
 because they showed that although the
typical interaction energy scale for electrons is on the order of
electron volts, only the binding interaction within pairs of
electrons, typically of O($10^{-3}eV$), need be included to obtain a 
quantitative theory of the
superconducting state. The underlying principle which makes this
reduction possible is the resilience of the Fermi surface to short-range
 repulsive
interactions.
As shown by Polchinski\cite{polchinski}, Shankar\cite{shankar}, and
others\cite{others}, all renormalizations from short-range repulsive
interactions are towards the Fermi surface.  As a result,
such interactions can effectively be integrated out leaving
behind dressed electrons or quasiparticles, thereby justifying the key
Landau tenet\cite{landau} that the
low-energy electronic excitation spectrum of a metal is identical to that of a
non-interacting Fermi gas.  Pairing is the only
wild card that destroys this picture. Since pairing instabilities
abound in metals for any number of reasons, for example,
Kohn-Luttinger anomalies\cite{kohn}, the Fermi surface is pure mathematical
fiction\cite{laughlinbook} at $T=0$. Nonetheless, our understanding of superconductivity
in metals would not be possible without it.  In this sense, superconductivity within the BCS account is subservient to the normal metallic
state as
superconductivity emerges as the unique interaction-driven instability of the
underlying electron Fermi surface.  

Essential then to the success of the BCS theory is a clean
identification of the natural propagating degrees of freedom in a
 system in which the short-range repulsive interactions may be of arbitrary
strength.   However, there are a number of experimentally relevant systems,
most notably the copper-oxide high-temperature superconductors, in
which such an identification of the propagating degrees of freedom in
the normal state has proved elusive, and as a consequence,
the nature of the superconducting state remains unresolved.  This
review focuses on the experimental and recent theoretical advances
which serve to
elucidate the nature of the propagating degrees of freedom in the
normal state of strongly correlated electron systems.  By strongly
correlated we mean systems in which no obvious
principle, such as that delineated\cite{polchinski,shankar,others} for Fermi liquids,
governs the renormalization of the electron-electron interactions.  A
ubiquity in such systems in which electron absorption is the
experimental probe is spectral weight transfer over large energy
scales, a phenomenon absent in Fermi liquids.  The presence of
UV-IR mixing is the tell-tale sign that electrons do not reside in electronic
states with well-defined energies.  Equivalently, the true propagating
modes are some admixture of multi-particle bare electron states. Precisely what is the nature of
the particles whose energies are sharp is the central question in
strongly correlated electron physics.  More precisely, what are the
particles for whom the single-particle Green function has poles with a
non-zero residue? Or equivalently, what is the natural low-energy
theory of a strongly correlated electron system?  We use natural here
to denote a theory in which there are no relevant perturbations.  Knowledge of such constituents
would permit a straightforward application of the BCS program because they render the
strongly correlated problem weakly interacting.  The
answer to this question ultimately resides in the physics of
collective phenomena. While collective phenomena can arise from states
of matter in which symmetries are broken, this need not be the case.
Noted examples include the Kondo effect in which all the electrons in
a metal act collectively to screen a local magnetic impurity.  Perhaps
the example which bares the closet resemblance to the physics here is that of quantum
chromodynamics (QCD).  In QCD, the pole in the single-quark propagator
vanishes at low energy and only bound quark states survive.
Precisely how such bound states are related to the degrees of freedom
in the UV is the hard problem of QCD. Our key message here is that
similar physics holds for doped Mott insulators.  Namely, composite
excitations emerge as the propagating degrees of freedom in doped Mott
insulators.  In the electron coordinates, these composite excitations
are able to explain the UV-IR mixing that is the fingerprint of the strong
correlations in doped Mott insulators as well as the anomalous
transport in the normal state of the copper-oxide superconductors.

\section{Mott's Problem: Insulating state of NiO}

\begin{figure}
\centering
\includegraphics[width=6.0cm]{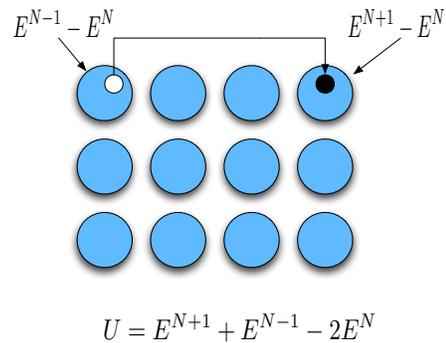}
\caption{A half-filled band as envisioned by Mott. Each blue circle
 represents a neutral atom with $N$ electrons and ground-state energy
 $E^{N}$.  The energy
 differences for electron removal and addition are explicitly shown.
Mott reasoned that no doubly occupied sites exist because at zero
temperature, $U=E^{N+1}+E^{N-1}-2E^{N}\gg 0$.  This is, of course, not
true. As a consequence the Mott gap must be thought of dynamically
rather than statically.}
\label{mott}
\end{figure}

The original Mott\cite{mott} problem grew out of coming to terms with why NiO
insulates.  Since NiO has two half-filled d-levels, it is
expected, based on the band picture of metals, to conduct at zero
temperature.  However, it insulates. For electrons to conduct, they
must hop from atom to atom.  An impediment to transport obtains if
the electron repulsions win out.  In Mott's construction, the relevant
interaction that dominates for narrow d-bands is the energy cost,
\beq
U=E^{N+1}+E^{N-1}-2E^N,
\eeq
for placing two electrons on the same Ni atom.  Here $E^N$ is the
ground state energy for an atom with $N$ valence electrons. Excluding
the filled levels, $N=2$ for Ni.  At zero
temperature, Mott reasoned that there is no Ni atom with $N\pm 1$
electrons if $U$ exceeds a critical value, typically on the 
order of the bandwidth.   In such a state, all Ni atoms have valence
of $+2$, and no conduction
obtains as illustrated in Fig. (\ref{mott}).  On this account, the
resultant charge gap is the energy cost for doubly occupying the same
site with spin up or spin down electrons.  
As a result, the simple Hamiltonian,
\beq\label{hubb}
H_{\rm Hubb}&=&-t\sum_{i,j,\sigma} g_{ij} c^\dagger_{i,\sigma}c_{j,\sigma}+U\sum_{i,\sigma} c^\dagger_{i,\uparrow}c^\dagger_{i,\downarrow}c_{i,\downarrow}c_{i,\uparrow},
\eeq
first introduced by Hubbard\cite{hubbard}, in which electrons hop among a set of
lattice
 sites, but pay an energy
cost $U$ whenever they doubly occupy the same site, is sufficient to
describe
 the transition to the state envisioned by Mott.  In this model,  $i,j$ label lattice sites, $g_{ij}$ is equal to one iff $i,j$
are nearest neighbours, $c_{i\sigma}$ annihilates an electron with
spin $\sigma$ on lattice site $i$ and $t$ is the nearest-neighbour
hopping matrix element.  In light of Eq. (\ref{hubb}), the simple Mott picture in which no Ni$
^{+++}$ or Ni$^+$ ions exist in the ground state, only works when the
hopping vanishes. This is the atomic limit.  In this extreme, the eigenstates of Eq. (\ref{hubb})
are indexed by the number of doubly occupied sites.  The propagating
modes are identified by bringing the interaction term,
\beq
H_U=U\sum_i
n_{i\uparrow}n_{i\downarrow}=\frac{U}{2}\sum_{i\sigma}\eta^\dagger_{i\sigma}\eta_{i\sigma},
\eeq
into quadratic form by defining $\eta_{i\sigma}= c_{i\sigma}^\dagger
n_{i-\sigma}$ which creates the excitations above the gap.  Its complement,
$\xi_{i\sigma}=c^\dagger_{i\sigma}(1-n_{i-\sigma})$ creates
  excitations strictly on empty sites and hence describes particle
  motion below the gap.  Consequently, in the atomic limit, the
  propagating degrees of freedom can be determined straightforwardly
  from the Hamiltonian.  

\begin{figure}
\centering
\includegraphics[width=8.0cm]{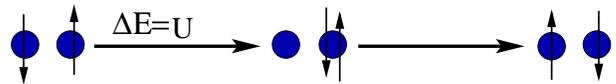}
\caption{Mechanism for the generation of the super-exchange
  interaction when the on-site interaction energy, $U$ exceeds the
  hybridisation energy, $t$.  The exchange energy will scale as $J=4t^2/U$.  }
\label{antif}
\end{figure}

Beyond the atomic limit, the propagating degrees of freedom responsible for the
gap are difficult to pinpoint primarily because even the lowest
eigenstate of the Hubbard model has doubly occupied character.  As a result, the charge gap in NiO
cannot be thought of in the terms envisioned by Mott, namely the gap to the first excited state
that has doubly occupied character.  Stated another way, the
decomposition of the electron operator as a sum of 
$\eta_{i\sigma}$ and $\xi_{i\sigma}$ is not canonical.  As a result,
$\eta$ and $\xi$ do not diagonalise the hopping term and hence do not propagate independently.  Precisely what the propagating degrees of freedom are that
are the efficient cause of the Mott gap is not known.  Even in the
case of one spatial dimension where the Hubbard model can be solved
exactly, it is not tractable to write down explicitly the band
structure of the degrees of freedom that become gapped\cite{1dhubb} at
strong coupling. The persistence of this problem led
  Laughlin\cite{laughlin} to assert that the Mott problem and all of
  its associated phenomena, such as the lower and upper Hubbard bands
  are entirely fictitious. He opts instead for antiferromagnetism as the
  cause of the gap in a half-filled band.  Indeed, antiferromagnetism
  and strong coupling Mott physics are closely related as illustrated
  in Fig. (\ref{antif}).  Electrons on neighbouring sites localized by large  on-site repulsions can exchange\cite{rvb} their spins if the spins are
  antiparallel.  Second order perturbation theory around the atomic
  limit is sufficient to establish that the energy scale for this
  process is $J\propto O(t^2/U)$ as illustrated in Fig. (\ref{antif}).
   As such processes lower the energy,
  long-range antiferromagentism is a natural consequence of correlation-induced
  localization of the electrons provided the lattice is, of course, bi-partite.

\begin{figure}
\centering
\includegraphics[width=9.0cm]{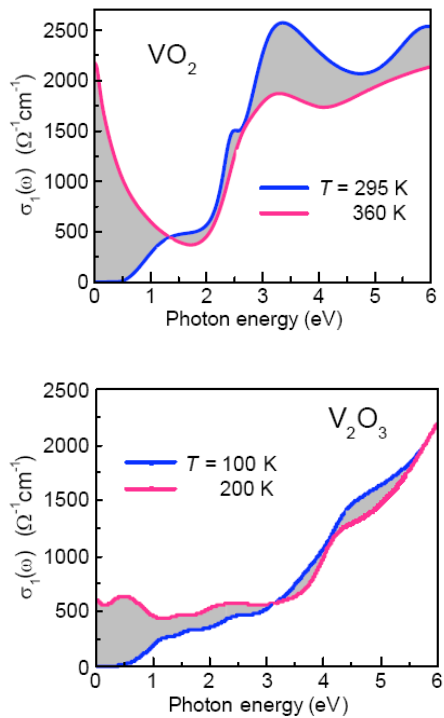}
\caption{Real part of the optical conductivity above and below the temperature for the
  onset of the Mott insulating state for 1) VO$_2$ and V$_2$O$_3$.  In both VO$_2$ and V$_2$O$_3$,
  the transition to the insulating state is accompanied by a transfer
  of spectral weight from in the vicinity of the chemical potential to
states as far as 6eV away.  This massive reshuffling of the spectral
weight upon the transition to the Mott state is a ubiquity of the Mott
transition. Data reprinted from PRB, {\bf 77}, 115121 (2008). }
\label{vo2}
\end{figure}

However, attributing the charge gap in transition metal oxides
 to symmetry-broken states such as antiferromagnets leaves an
explanatory residue.  It is that
residue that we term  Mottness\cite{mottness}.  Consider a
prototypical Mott system
 VO$_2$ which undergoes\cite{v1,v2,v3,v4,v5,v6,v7,v8,v9,lupi3}
a transition to an insulating state at roughly 340K. In this system,
each vanadium ion has a valence of $+4$ giving rise to a half-filled
$d^1$ configuration.  Below 340K, the
conductivity decreases by four orders of magnitude and a charge gap of
$0.6eV$ opens\cite{v6}.  While no hint of magnetic order is detected in this
system, the vanadium atoms do pair up to form tilted dimers along the
c-axis\cite{v1,v2,v3,v4,v5}.  Consequently, a similar
question\cite{v2,v8,v9,mottv} has arisen in this
system as to whether or not the gap is due to
symmetry breaking resulting in a doubling of the unit cell or
the correlation picture of Mott.  This question has persisted even
though Mott\cite{mottv} pointed out that a gap of 0.6eV is beyond any
energy scale entailed by
dimerization of the vanadium ions. 
The optical response of this system probed
by ellipsometry is particularly useful here in settling this question.  The key feature shown in
Fig. (\ref{vo2}) is that lowering the temperature to 295K\cite{v6}, an
energy scale considerably less than $0.6eV$,
leads to transfer of spectral weight from states in the
vicinity of the chemical potential (red region) to those at considerably high
energies (blue region), roughly 6 eV away and beyond.  Such energy
scales over which the spectral weight is redistributed vastly exceed
those relevant to dimerization.  They are, however, consistent with correlation physics on the
 $U$-scale. This is Mottness and it persists, as we will see, even
 upon a transition to the superconducting state.  In fact, this state of affairs obtains even in
 half-filled bands, for example, V$_2$O$_3$, which are
 known\cite{v6,v2031,v2032,v2033,v2034,lupi2} to order
 antiferromagnetically in the insulating state.  The second panel in
 Fig. (\ref{vo2}) shows the analogous optical conductivity across the
 metal-to-insulator transition in V$_2$O$_3$.  As in the case of
 VO$_2$, the transition to the insulating state also involves transfer of
 spectral weight from the chemical potential to states at least $6eV$
 away.  This mixing of high and low energy scales is a ubiquity in
 Mott systems.  Hence, central to the charge gap are propagating degrees
of freedom that entail the $U$ scale, not simply the smaller energy
scale associated with whatever ordering phenomenon might obtain. 

\begin{figure}
\centering
\includegraphics[width=9.0cm]{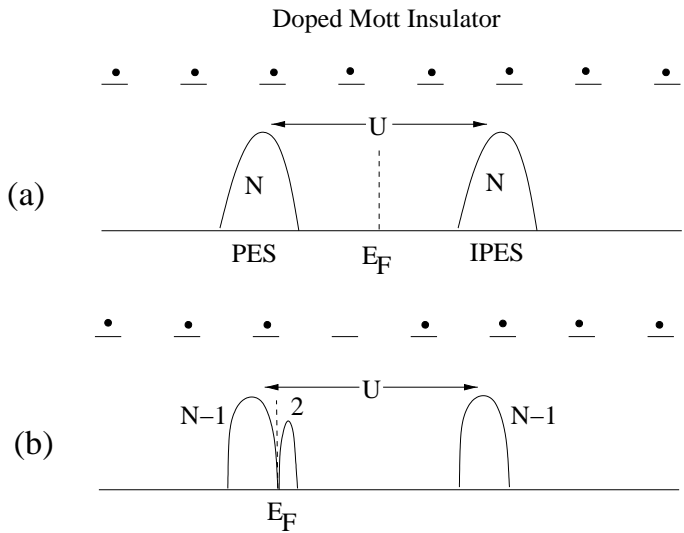}
\caption{a.) Half-filled band in the limit of large $U\gg t$ results in a
 gap in the single particle spectrum.  The bands shown represent the
 upper and lower Hubbard bands. In an $N$-site system, the total
 number of single-particle states in each band at half-filling is $N$.
  The electron-removal band, the photo-emission  (PES) and
  electron -addition bands, inverse-photoemission (IPES) bands
  correspond to an electron moving on
 empty and singly-occupied sites, respectively. In the atomic limit
 the splitting between the bands is $U$.  
b.) Evolution of the single-particle density of states from
  half-filling to the one-hole limit in a doped Mott insulator
  in the atomic limit of the Hubbard model.  Removal of an electron results
  in two empty states at low energy as opposed to one in the
  band-insulator limit. The key difference with the Fermi liquid is
  that the total weight spectral weight carried by the lower Hubbard
  band (analogue of the valence band in a Fermi liquid) is not a
  constant but a function of the filling. }
\label{fig2}
\end{figure}

UV-IR mixing, as evidenced by the spectral weight transfer in the
optical conductivity, indicates that any single-electron description of a Mott
system is moot.  In fact,  Mott insulators are characterized\cite{zeros1,zeros2,zeros3,myzeros} by a
vanishing of the single-particle electron Green function along a connected
surface in momentum space.  While the volume of this zero surface is not
directly tied to the particle density except in the case of
particle-hole symmetry\cite{zeros3,myzeros}, in direct contrast to the
surface enclosed by the divergence of the Green function in a Fermi
liquid, the zero surface for a Mott insulator, nonetheless, has a profound
significance.  It indicates that electrons are not the
propagating degrees of freedom that give rise to the gapped
spectrum.  A correct identification of the
propagating modes would result in a pole in the associated
single-particle Green function.  An analogy to QCD is in order here.
At IR energy scales, the single-quark propagator vanishes.  However,
the meson or bound quark propagator has a pole. The Mott problem
amounts to finding the particle whose pole in the single-particle
propagator leads to the band structure of a Mott insulator depicted in
Fig. (\ref{fig2}).  The only alternative is that
some sort of dynamically generated composite or bound state accounts for the gap in the
spectrum.  That the natural propagating modes
responsible for the gapped structure of a Mott insulator are composite
particles or bound states of the elemental excitations can be seen
from a simple physical argument.  Beyond the atomic limit of a Mott
insulator, double occupancy explicitly occurs in the ground
state. Double occupancy cannot occur without the simultaneous creation
of empty sites. If the empty sites move freely, then the Mott insulator is
in actuality a conductor.  As is evident from Fig. (\ref{antif}),
mobile double occupancy will also destroy local antiferromagnetic
correlations.  Hence, both the magnetic and electrical properties of a
Mott insulator demand that doubly occupied (doublon) and empty sites (holon) form bound
states. In fact, Mott\cite{mott} anticipated as much in so far as the
insulating properties are concerned.  Doublon-holon
binding has been observed to lower the energy in variational
approaches to the half-filled Mott band\cite{fulde,numerical}. In a
similar vein, Castellani and colleagues\cite{castellani}
argued that the Mott state is one in which double occupancy is
localized but delocalized in the metal.  While there is a
tendency in the literature\cite{gros} to regard doubly occupied and
empty-site
bound states as simply virtual double occupancy, this is a
misnomer. The number of doubly occupied sites in the Mott state is
finite\cite{castellani,fulde} in contrast to the Brinkman-Rice mechanism in which it vanishes
identically\cite{brice}.  Further, as can be seen by diagonalising a
small Hubbard cluster, not all doubly occupied sites in a
half-filled band occur because they mediate $J-$ scale physics.
Consequently, facing up to the Mott problem away from the atomic limit
requires an explicit mechanism for the localization of double occupancy.
  Such a
mechanism underlies the propagating degrees of freedom which are
responsible for the gap and ultimately the onset of antiferromagnetic
order.  While it is tempting to invert the problem and invoke antiferromagnetism as the
mechanism for the localization of double occupancy, this is
problematic because it leaves Mottness unexplained as noted previously\cite{anderson}.  We offer here an
explicit construction of the propagating degrees of freedom underlying
the Mott gap. 

\noindent

\section{Doped Mott Insulators: Breakdown of Fermi liquid theory}

\begin{figure}
\centering
\includegraphics[width=6.0cm]{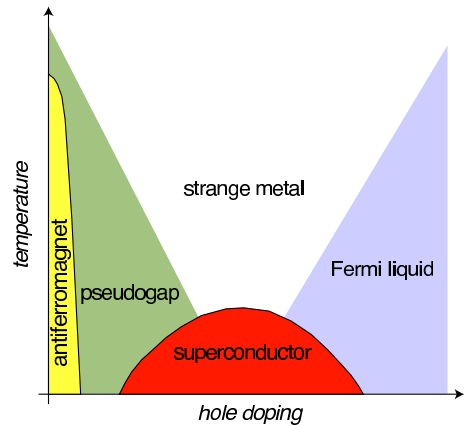}
\caption{Heuristic phase diagram of the copper-oxide
  superconductors.  In the
 strange metal, the resistivity is a linear function of
 temperature. In the pseudogap the single-particle density of states
 is suppressed without the onset of global phase coherence indicative
 of superconductivity.  The dome-shape of the superconducting region
 with an optimal doping level of $x_{\rm opt}\approx 0.17$ is quantitatively
accurate only for La$_{2-x}$Sr$_x$CuO$_4$.  }
\label{pdiagram}
\end{figure}

The normal state of the cuprates embodies a panoply of phenomena that
are inconsistent with Fermi liquid theory.  The most vexing
 are the strange metal, characterised by $T-$ linear resistivity\cite{ando,raffy}, as
 opposed to the quadratic
dependence predicted in Fermi liquids, and the
pseudogap\cite{pg1,pg2,pg3} in which the single-particle density of states is suppressed, although the
superconducting gap vanishes.  We highlight these features here
because the phase diagram of the cuprates, Fig. (\ref{pdiagram}),
indicates unambiguously that the correct theory of the superconducting
state must at higher temperatures account for a charge vacuum (that is the electronic state)  that is capable of
explaining how the onset\cite{ando,raffy} of the pseudogap at the temperature scale $T^\ast$ leads to a cessation of
T-linear resistivity.   While
numerous theories of the pseudogap
abound\cite{stripes1,ddw,rvb,inco1,inco2,
inco3}, none offer a resolution of
the $T-$linear resistivity problem within any realistic model of a
doped Mott insulator.  Part of the problem is that a series of
associated phenomena, for example, incipient diamagnetism\cite{nernst}
indicative of
incoherent pairing\cite{inco1,inco2,inco3}, electronic
inhomogeneity\cite{stripes1,stripes2,stripes3,stripes4,stripes5,stripes6}, time-reversal
symmetry breaking\cite{trsb1,trsb2,trsb3,trsb4}, and quantum oscillations\cite{qoscill} in
the Hall conductivity,
possibly associated with the emergence of closed electron (not
hole) pockets in the first Brillouin zone (FBZ), obscure the
efficient cause of the pseudogap and its continuity with the strange
metal. What we propose here is that the degrees of freedom responsible
for dynamical spectral weight transfer are directly responsible for
the pseudogap and the transition to the strange metal.

\subsection{Dynamical Spectral Weight Transfer:  More than just electrons}

In the electronic state or charge vacuum that accounts for the normal
state of the cuprates, the key assumption of Fermi
liquid theory that the low-energy spectra of the interacting and free
systems bare a one-to-one correspondence must break down.   More precisely, the
interacting system must contain electronic states at low energy that
have no counterpart in the non-interacting system.  One
possibility\cite{ruckenstein} is that spectral weight transfer between high and low
energies mediates new electronic states at low energy that have no counterpart in the non-interacting system, thereby
leading to a breakdown of Fermi liquid theory. We show in this section
that this is precisely what obtains in the Hubbard model.  

\begin{figure}
\centering
\includegraphics[width=6.0cm]{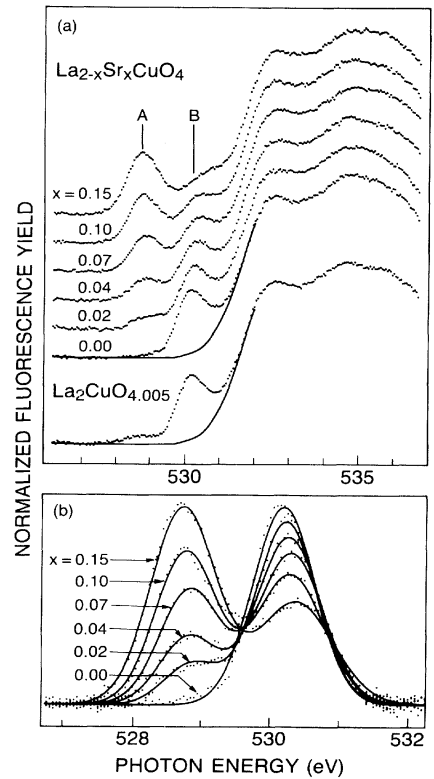}
\caption{a) Normalized fluorescence\cite{chen} yield at the oxygen K edge of
 La$_{2-x}$Sr$_x$CuO$_{4+\delta}$.  In the undoped sample, the only
 absorption occurs at 530eV, indicated by B.  Upon doping the
 intensity at B is transferred to the feature at A, located at
 528eV.  b) Gaussian fits to the absorption features at A and B with
 the background subtracted. Reprinted from Phys. Rev. Lett. {\bf 66}, 104 (1991).}
\label{fy}
\end{figure}

To
motivate this pathway for the breakdown of Fermi liquid theory, we
analyse the Oxygen 1s x-ray absorption
experiments\cite{chen}  on
La$_{2-x}$Sr$_x$CuO$_4$ (LSCO) shown in
Fig. (\ref{fy}).  In such
experiments, an electron is promoted from the core 1s to an
unoccupied level.  The experimental observable is the fluorescence
yield as a function of energy as electrons relax back to the
valence states. The experiments, Fig. (\ref{fy}), show that at
$x=0$, all the available states lie at 530eV.  As a function of
doping, the intensity in the high-energy peak decreases and is
transferred to states at 528eV.  In fact, experimentally, the lower peak grows faster
than $2x$ while the upper peak decreases faster than $1-x$, $x$ the
number of holes.  In La$_{2-x}$Sr$_x$CuO$_4$, the doping level $x$ can
be unambiguously determined because each Sr atom produces one hole. The separation between these two
peaks is the optical gap in the parent insulating material.  

\begin{figure}
\centering
\includegraphics[width=8.0cm,]{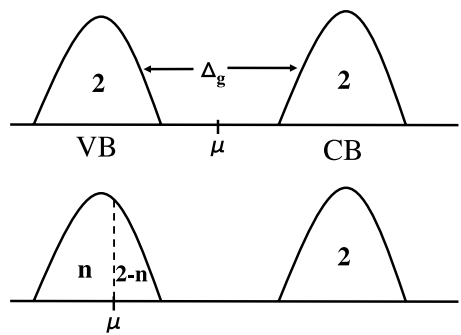}
\caption{Evolution of the single-particle density of states in a Fermi
  liquid in the valence band as a function of the electron filling, $n$.  The total weight of the valence band is a constant 2, that
  is, 2 states per site.  Doping simply pushes one state above the
  chemical potential.  The integral of the density of states below the
chemical potential is always the filling, $n$.}
\label{fig1}
\end{figure}

The redistribution of the spectral weight seen in the experiments can
be explained within the Hubbard model. Since the experiments are
probing the flourescence yield into the available low-energy states,
the relevant theoretial quantity is the number of single-particle
 addition states per site at low energy,
\beq\label{dossum}
L=\int_\mu^\Lambda N(\omega)d\omega,
\eeq
defined as the integral of the single-particle density of states
($N(\omega)$) from the chemical potential, $\mu$, to a cutoff energy
scale, $\Lambda$, demarcating the division between the IR and UV scales. In a Fermi liquid, $\Lambda$ can be
extended to infinity as there is no upper band, whereas in a semiconductor,
$\Lambda$ should extend only to the top of the valence band to count
the states available upon the addition of holes.  To calibrate this
quantity, we compare it with the number of ways electrons can be added
to the empty states created by the dopants. Let this quantity be
$n_h$.   Consider first
the case of a Fermi liquid or non-interacting system.  As illustrated
in Fig. (\ref{fig1}), the total weight of the valence band is 2, that
is, there are 2 states per site. 
\noindent The integrated weight of the valence
band up to the chemical potential determines the filling.
Consequently, the unoccupied part of the spectrum, which determines $L$,
is given by $L=2-n$. The number of ways electrons can be added to the
empty sites is also $n_h=2-n$ (see Fig. (\ref{fig1})).  Consequently, the number
of low-energy states per electron per spin is identically unity. The
key fact on which this result hinges is that the total weight of the
valence band is a constant independent of the electron density.

By
contrast, a doped Mott insulator behaves quite differently.  At half-filling the chemical potential
lies in the gap as depicted in Fig. (\ref{fig2}).  The sum rule that 2 states exist per site 
applies only to the combined weight of both bands.  At any finite
doping,
the weight in the LHB and UHB is determined by the density.  As a
consequence, there is no independent sum rule for the occupied and
empty parts of each band.  That is, there is no independent sum rule
for $L$.  Consider first the atomic limit.  In this limit, the total
spectral weight of the lower band,
\beq
{\rm m}^0_{\rm LHB}
	= \frac{1}{N}\sum_{i,\sigma} \langle\{\xi_{i\sigma},\xi_{i\sigma}^\dagger\}\rangle=2-n,
\eeq
is given by the anticommutator of the operators that create and
annihilate
singly occcupied sites.  Since 
each hole in a half-filled band decreases the double occupancy by one, the weight of the UHB is $1-x$.  Because the total weight of the
UHB and LHB must be 2, we find that $2-n+1-x=2$ or $n=1-x$ and ${\rm
  m}^0_{\rm LHB}=1+x$ in the atomic limit. The weights $1+x$ and $1-x$ also determine 
the total ways electrons can occupy each of the bands.  Thus, in the atomic limit, electrons 
alone exhaust the total degrees of freedom of each band. Further,
since each hole 
leaves behind an empty site that can be occupied by either a spin up or a spin 
down electron, the electron addition spectrum in the LHB has weight 
$L=2x$\cite{sawatzky}.  Hence, the occupied part of the LHB and UHB both 
have identical weights of $1-x$ in the atomic limit as depicted in
Fig.  (\ref{fig2}).  

\begin{figure}
\includegraphics[width = 8.0cm]{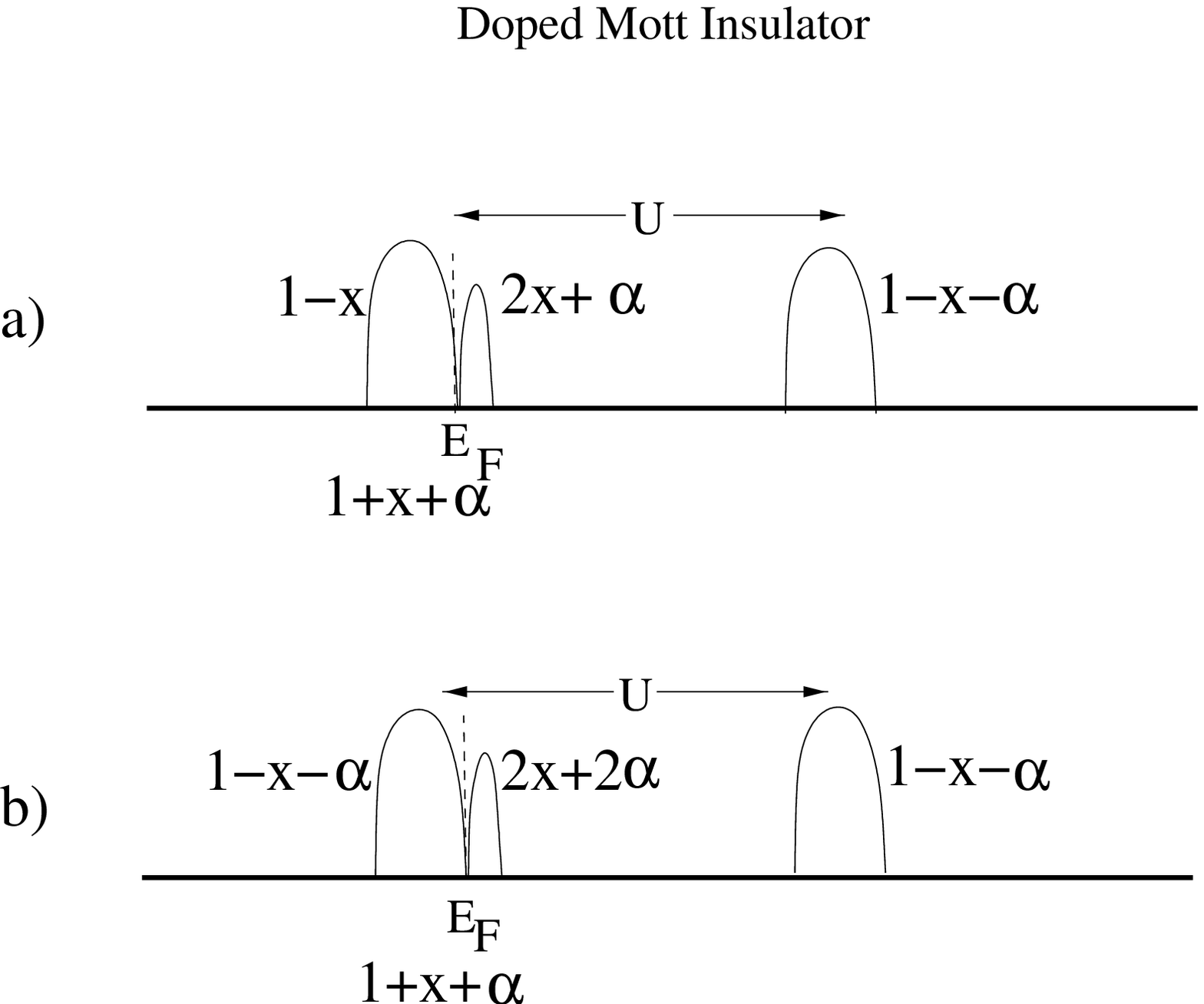}
\caption{Redistribution of spectral weight in the Hubbard model upon doping the
		insulating state with $x$ holes. $\alpha$ is the dynamical correction 
		mediated by the doubly occupied sector. To order $t/U$, this correction 
		worked out by Harris and Lange\cite{harris}. a) The traditional approach
		\cite{sawatzky,stechel} in which the occupied part of the 
		lower band is fixed to the electron filling $1-x$.  b)
                New assignment of the spectral weight in terms of
                dynamically generated charge carriers.  In this
                picture, the weight of the empty part of the LHB per spin is
                the effective doping level, $x'=x+\alpha$. }
\label{spec}
\end{figure}

Explaining the experiments fully necessitates going beyond the atomic limit. Away from this limit, the total spectral weight of the LHB
\beq\label{hl}
{\rm m}_{\rm LHB}
	=1+x+\frac{2t}{U}\sum_{ij\sigma}g_{ij}
		\langle f_{i\sigma}^\dagger f_{j\sigma}\rangle+\cdots
	=1+x+\alpha,
\eeq
has $t/U$ corrections\cite{harris} which are entirely positive.  Here 
  $f_{i\sigma}$ are related to the original bare fermion operators via a 
canonical transformation that brings the Hubbard model into block diagonal 
form in which the energy of each block is $nU$. In fact, all orders of 
perturbation theory\cite{eskes,harris} increase the intensity of the LHB beyond its 
atomic limit of $1+x$. It is these dynamical corrections that $\alpha$ 
denotes. That
$\alpha$ is positive can be seen from the simple fact that turning on
the hopping increases the total weight of the LHB from the
atomic limit of one per site to ultimately two per site in the non-interacting
limit. This increase beyond the atomic limit of the intensity of
the LHB is significant because the number of ways of assigning
electrons to the LHB remains fixed at $1+x$.  As a result, electrons
alone do not exhaust the degrees of freedom in the LHB, in direct
contrast to the atomic limit. That is, a low-energy theory of the LHB
requires new non-fermionic charge degrees of freedom. Nonetheless,
there is a conserved charge given simply by the electron filling.
Since the low-energy theory must have both fermionic and non-fermionic
degrees of freedom, there are clearly less fermionic quasiparticles
than there are bare electrons. Consequently, the Landau Fermi liquid one-to-one correspondence
between the two fails.  We propose that the chemical potial for the
effective number
of
low-energy fermionic
degrees of freedom can be determined by partitioning the
spectrum in the LHB so that dynamical spectral weight transfer is
essentially removed.  In such a picture, the empty part of the
spectrum per spin is equal to the weight removed from the occupied
part of the LHB when a hole is created.
 Hence, we arrive at the assignments of the spectral weights in
Fig. (\ref{spec}b) in which the doping level is renormalized by the
dynamics; that is, $x'=x+\alpha$.  In other words, the 
dynamical degrees of freedom denoted by $\alpha$ serve to supplement the effective 
phase  space of a hole-doped system and $x'=x+\alpha$ now denotes the effective 
number of hole degrees of freedom per spin at low
energy.  Any experiment that couples to  the fermionic low-energy degrees of 
freedom should be interpreted in terms of the total number of hole degrees of freedom, 
$x+\alpha$ not $x$.  In terms of the true fermionic quasiparticles,
$L=2x'$ which should be compared with $n_h=2x$.  Clearly, $L/n_h>1$
(as is seen experimentally)
and Fermi liquid theory fails.  This failure\cite{dswtfinal} arises ultimately because
the chemical potential for the fermionic degrees of freedom which make
the doped Mott system weakly interacting is less than that of the bare
electrons. Precisely how Fermi liquid theory re-emerges beyond a
critical doping is detailed elsewhere\cite{dswtfinal}.

Experiments on the temperature dependence of the Hall coefficient
offer direct confirmation that the doping level is renormalized
dynamically.  In LSCO, the
Hall number has been fit~\cite{gorkov} to a
function of the form,
\beq\label{deltac}
n_{\rm Hall}(T,x)=n_0(x)+n_1(x)\exp[-\Delta(x)/T]
\eeq
where $n_0(x)$ is temperature independent. Empirically\cite{gorkov}
$\Delta(x)$ is related to the pseudogap scale.   Consequently, there
seem to be two types of charges in LSCO, ones that arise from the
doping and others which emerge from the transfer of anti-bonding
states down to the chemical potential producing effectively bound
charge states.  Our key contention here is that $\alpha$ arises from
the second term in Eq. (\ref{deltac}).  In part, it is the presence of these two distinct kinds of charge
carriers, inferred from the Hall number, that has motivated a two-fluid
model of the cuprates\cite{gorkov,pines}.   One of the key points of this Colloquium is that
two types of charges are already implied by $L>2x$ as noted in Fig. (\ref{spec}b).  The extra
degrees of freedom arise from the hybridization with the doubly occupied
sector. Interestingly, the same hybridization arises even in the
half-filled band. However, a gap appears.  Accounting for this
difference is a charge 2e boson which forms bound states with a hard
gap at half-filling but only a pseudogap in the doped case as implied by Eq. (\ref{deltac}).  This new degree of freedom emerges only when the
high-energy sector is integrated out exactly.  In fact, this procedure proves quite
generally that integrating out the high energy scale in a doped Mott
insulator and in a Fermi liquid is fundamentally different.  In the
latter, no new degrees of freedom are generated, whereas in the former
a new charge 2e boson emerges\cite{charge2e,charge2e1,charge2e2,charge2e3,charge2e4}.  The charge 2e excitation mediates new electronic states at low
energies by binding to a hole and hence explains naturally the
temperature-dependent component of the Hall number.  Such bound states generate
 a pseudogap and their unbinding yields $T-$linear resistivity. 

\subsection{Optical Conductivity: Emergence of the pseudogap}

\begin{figure}
\centering
\includegraphics[width=7.0cm]{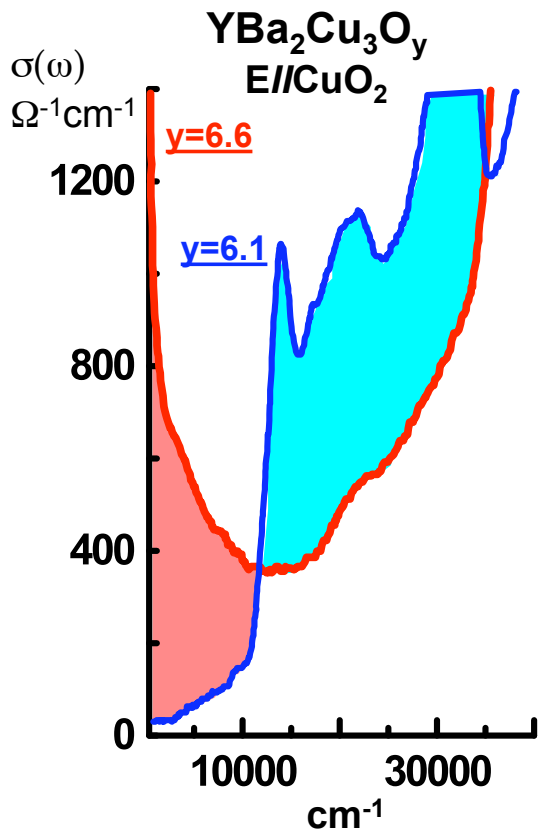}
\caption{Evolution of the in-plane optical conductivity in
  YBa$_2$Cu$_3$O$_y$ (YBCO) for two doping levels.  To a good facsimile,
  $y=6.1$ represents the parent material and $y=6.6$, an overdoped
  sample. The key feature is the transfer of spectral weight above
  $1.2eV$ in the `parent' material to states in the vicinity of the
  chemical potential upon doping. Reprinted from PRB, ${\bf 47}$, 8233 (1993).  }
\label{opttrans}
\end{figure}

\begin{figure}
\centering
\includegraphics[width=7.0cm,angle=90]{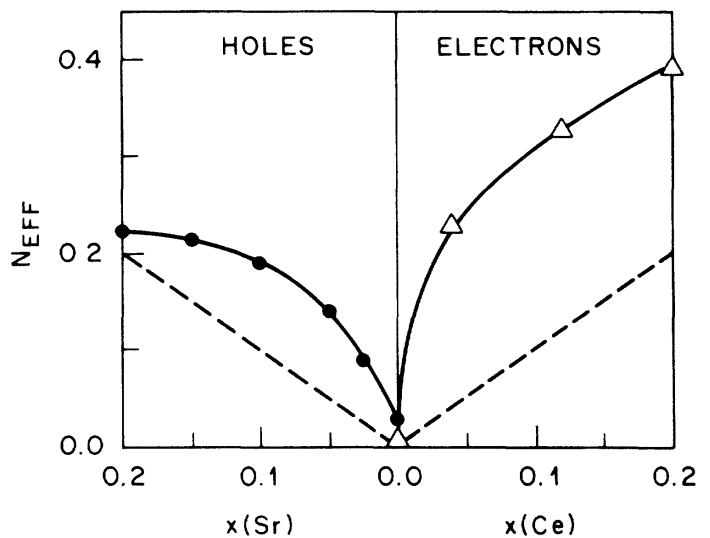}
\caption{Integrated optical conductivity for a hole-doped (YBCO) and
an electron-doped (Nd$_{2-x}$Ce$_2$CuO$_4$ (NCCO)) cuprate reflecting the
  normalized carrier density.  The dashed line
  indicates what is expected for a doping a semiconductor. Within the
  Hubbard model, the charge density in excess of the nominal doping
  level is generated by dynamical spectral weight transfer. Reprinted
  from PRB, {\bf 41}, 11605 (1990). }
\label{neff}
\end{figure}

Optical conductivity experiments also indicate that the effective number of
charge carriers exceeds the nominal count provided by the doping as
predicted by the quasiparticle picture in Fig. (\ref{spec}b).  To
see this, we compute the effective number of
carriers, or more precisely the normalized carrier density, by integrating the
optical conductivity 
\beq
N_{\rm eff}(\Omega)=\frac{2m V_{\rm cell}}{\pi e^2}\int_0^\Omega
\sigma(\omega)d\omega
\eeq
up to the optical gap $\Omega\approx 1.2eV$. Here $\sigma(\omega)$ is the optical conductivity, $V_{\rm cell}$ the
unit-cell volume per  formula unit, $m$
the free electron mass, and $e$ the electron charge.  In a rigid-band
semiconductor model in which spectral weight transfer is absent,
$N_{\rm eff}=n_h$.  Shown in Fig. (\ref{opttrans}) is the optical
conductivity\cite{cooper,basovopt1,lupi1} for YBa$_2$Cu$_3$O$_y$ along the CuO$_2$ plane for
oxygen doping levels of $y=6.1$ and $=6.6$.  At $y=6.6$, the strong
suppression of the optical conductivity below $\approx 1.2eV$ vanishes
and is accompanied by a decrease in the
spectral intensity in the high-energy sector. This behaviour is
analogous to the inversion of the spectral weights above 340K in the
VO$_2$ upon a transition to the metallic state.  In fact, the energy
scales in the doping-induced metallic state are identical to that in
the temperature-induced transition in these two systems.  To quantify
the transfer of spectral weight, we plot the effective number\cite{coopernum} of
carriers $N_{\rm eff}$ (shaded red region) that fill-in the optical gap in the `parent'
material as a result of doping.  As is evident from Fig. (\ref{neff}),
the normalized carrier
density exceeds the carrier density that would obtain in a doped
semiconductor model (the dashed line). This result is significant
because independently it was shown\cite{basovopt1} that throughout the
underdoped regime of the cuprates, the effective mass is constant.  As
a result, the Mott transition proceeds by a vanishing of the carrier
number rather than the mass divergence of the Brinkman-Rice
scenario\cite{brice}. Consequently, the strong deviation from $2x$
is a clear indicator that dynamical spectral weight transfer is
operative as well in the optical conductivity.  This is expected since
the current couples to the true low-energy degrees of freedom and
hence the quasiparticles identified in Fig. (\ref{spec}b) which have an
effective doping level of $x'=x+\alpha>x$ are
relevant.   A final feature which
deserves mention is the non-zero intercept of $N_{\rm eff}$ at $x=0$.
This suggests that even at half-filling, a remnant of the excitations
that fill-in the spectral density below the Mott gap is present.  

\noindent
\noindent

\begin{figure}
\centering
\includegraphics[width=9.0cm]{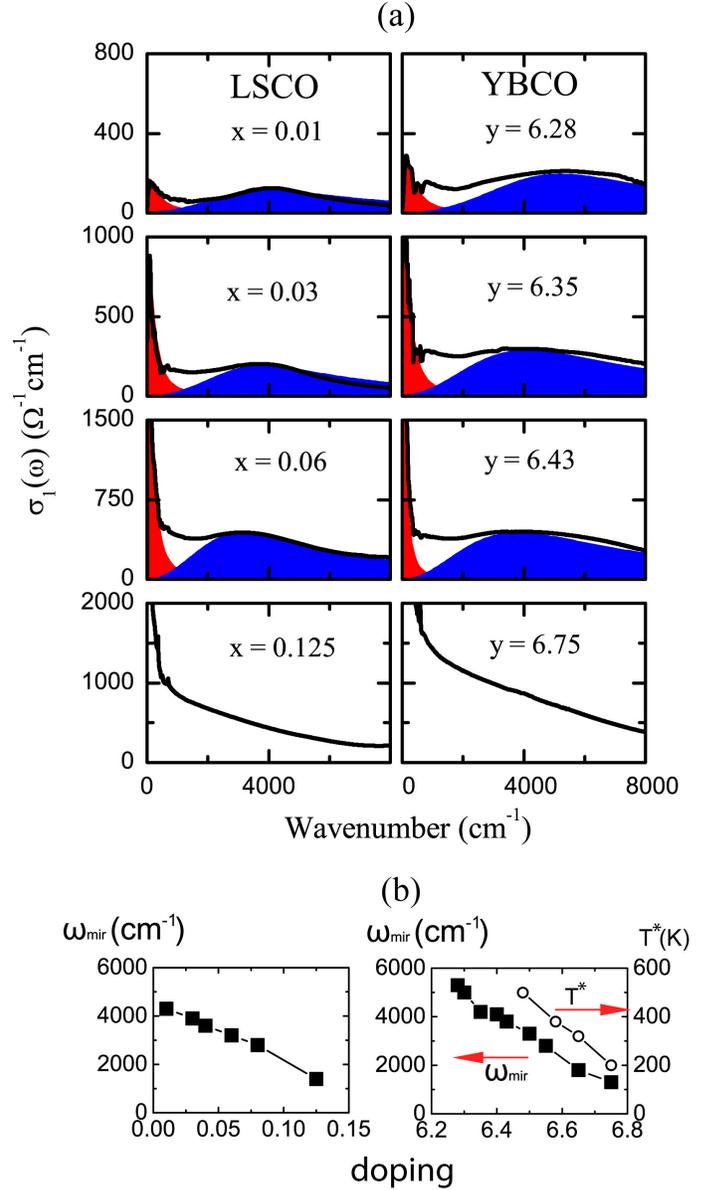}
\caption{a) Optical conductivity for LSCO (left panel) and YBCO (right
  panel) as a function of
  frequency in the underdoped regime.  The top two contain
  measurements of the optical conductivity at 10K for
  nonsuperconducting samples, the next four show $\sigma(\omega)$ at $T=T_c$ for
  superconducting samples, whereas in the
  bottom panel, the temperature is roughly $T^*$ for the $y=6.75$
  material.  In all spectra, phonons have been removed by fitting them
  with Lorentzian oscillators.  Clearly shown in the underdoped
  samples is a Drude-like peak (red) followed by a resonance in the
  mid-infrared (blue).  The appearance of the mid-IR below $T^*$
  indicates a strong connection with the pseudogap. b)  Frequency of
  mid-infrared, $\omega_{\rm MIR}$, and the pseudogap onset
  temperature $T^*$ as a function of doping.  The combined onset of the
mid-IR below $T^*$ and the fact that $\omega_{\rm MIR}$ and $T*$ have
the same doping dependence indicates that they share a common
cause. Reprinted from, PRB, {\bf 72 }, 054529 (2005).  }
\label{optcond2}
\end{figure}

The precise nature of the excitations that lead to a filling-in of the
spectral weight immediately above the lower Hubbard band is intimately
tied to the pseudogap.  As such states emerge from high energy, they arise from
the incoherent part of the spectrum and hence must necessarily be
anti-bonding and hence non-current carrying. That is, they describe
localized excitations.  This is consistent with the fact that the
operators which generate\cite{eskes} the dynamical part of the spectral weight are
inherently local.  Alternatively, $L>2x$ means that while there are $L$ ways to
add a particle, there are only $2x$ ways to add an electron.  This
mismatch means that some states must be orthogonal to the addition of
an electron.  Consequently,
the spectral function at certain momenta contains states in which the
spectral weight vanishes arbitrarily close to the
chemical potential. For $L-n_h$ of the particle addition states, a
quasiparticle peak is absent in the electron addition spectrum.  The result is a pseudogap as all k-states are
not necessarily gapped.  Although it has
been known for some time\cite{harris} that $L>2x$ at strong coupling in a doped
Mott insulator based on the Hubbard model, that this simple fact implies a pseudogap has
not been deduced previously.  This conclusion is borne out
experimentally. While the pseudogap is most easily seen through the c-axis
optical conductivity, it can also be detected, though more subtlety,
from in-plane transport measurements.  As the lower panel in Fig. (\ref{optcond2}a) indicates\cite{basovopt1},
the optical conductivity displays broad features at intermediate
energies with a well-defined peak at zero frequency.  This peak is
well-described by a $1/\omega^2$ dependence in a Drude model.
However, below $T^*$, a distinctive band appears\cite{basovopt1} in the optical
conductivity in the mid-infrared energy range.  The central frequency of this mid-infrared feature
evolves to lower energies and gradually disappears in heavily
overdoped samples.  The evolution\cite{basovopt1} of $\omega_{\rm MIR}$ as a function
of doping (see Fig. (\ref{optcond2}b)
tracks well that of $T^*$, thereby corroborating that the mid-IR
feature and the pseudogap are related phenomena.  While the origin of
the mid-IR remains a point of controversy\cite{shilaoc,millisoc,midir3,hkoc,midir5} in cuprate phenomenology,
two things are clear.  First, any explanation of it must apply equally
to the pseudogap regime.  Second, the correct explanation must involve
dynamical spectral weight transfer from the upper Hubbard band as it
is from this band that the mid-IR spectral intensity originates.  The
latter implies that the mid-IR should be absent from low-energy
reductions of the Hubbard model which ignore double occupancy.  It is
for this reason that a recent analysis\cite{midir} of the mid-IR within the t-J model
has found it necessary to invoke phonons to account for this
resonance. In fact, Uchida, et al.\cite{uchida} were the first to point out that
because the mid-IR scales as $t$, such physics is beyond that captured
by the t-J
model, at least in its traditional implementation in which the
operators are not transformed (see first Appendix).   However, in the Hubbard model, be it
the single-band\cite{shilaoc,millisoc,lupi5,lupi4}  or the multi-band Hubbard
model\cite{hkoc}, despite the differences
in the computational scheme, a mid-IR resonance appears.  Within the computational
scheme used by Chakraborty, et al.\cite{shilaoc}, it is possible to
determine which local correlations on a plaquette contribute to the
mid-IR.  They found that of the 256 plaquette states, when just a
single state was eliminated, the mid-IR feature vanished.  Key
features of this state are that 1) it contains a mixture of singly
 (87$\%$) and doubly occupied (13$\%$) sites, 2) its
energy is $-1.3t$, essentially the energy of mid-IR peak, and 3) the
spatial symmetry of the eigenstate is d$_{x^2-y^2}$.   The latter is
particularly revealing because it indicates that the mid-IR is highly
anisotropic and has the same momentum dependence as does the
pseudogap, thereby corroborating the experimental\cite{basovopt1} trend in
Fig. (\ref{optcond2}b) that the pseudogap and mid-IR share the same origin.

\noindent

\subsection{Scale Invariance: Beyond one parameter}

\begin{figure}
\centering
\includegraphics[width=7.0cm,angle=90]{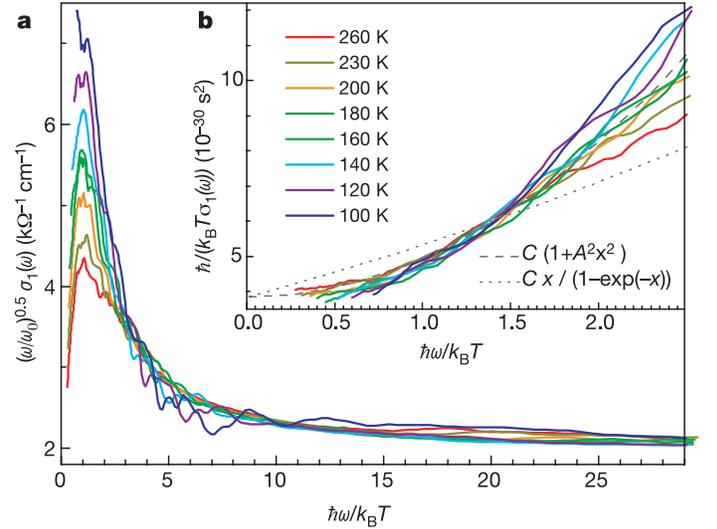}
\caption{Temperature/frequency scaling behaviour of the real part of
  the optical conductivity of
  Bi$_2$Sr$_2$Ca$_{0.92}$Y$_{0.08}$Cu$_2$O$_{8+\delta}$ (BSCO).   The data
are plotted as a function of
$(\omega/\omega_0^{0.5}\sigma(\omega,T))$.  Using a function of the
form,
$T^{-\mu} f(\omega/T)$ , we observe a) collapse of all curves for
$\mu=0.5$ for $\hbar\omega>k_B T$ and b) collapse for $\mu=1$ for
$\hbar\omega/k_BT<1.5$.  Such a change in the exponent $\mu$ is not
consistent with the general assumption of scale invariance. Reprinted
from Science, {\bf 425}, 271 (2003). }
\label{optscal}
\end{figure}

Scale invariance is a fundamental property of critical matter.  For
quantum matter\cite{hertz} that is critical, scale invariance implies that the system looks the
same on average at any chosen time and length
scale.  Provided that $\omega<T$, quantum critical systems exhibit
classical  behaviour in which only the temperature governs the
dynamics in the vicinity of the critical point.  In this quantum critical regime, as it is called, the
transport relaxation rate between the quasiparticles is universal in
that in can be deduced,
\beq\label{tr}
\frac{1}{\tau_{\rm tr}}=\frac{k_B T}{\hbar},
\eeq
 by using dimensional analysis on
Boltzmann's constant, $T$ and $\hbar$, Planck's constant.  Coupled with the assumption of scale invariance, we posit as
well a
Drude form for the optical conductivity,
\begin{eqnarray}
\label{drude}
\sigma_{\rm Drude}=\frac{1}{4\pi}\frac{\omega^2_{\rm pl}\tau_{\rm
    tr}}{1+\omega^2\tau^2_{\rm tr}}, 
\end{eqnarray} 
with $\omega_{\rm pl}$ the plasma frequency.  These two assumptions
lead immediately to
$T-$linear resistivity when
Eq. (\ref{tr}) is substituted into Eq. (\ref{drude}) and the
zero-frequency limit is taken.  It is for this reason that $T-$ linear
resistivity is commonly attributed to quantum criticality.  Two
questions are relevant here, however, to determine if this procedure
is internally consistent with the underlying hypothesis of scale invariance:
1) Does Eq. (\ref{drude}) describe the cuprates in any doping range,
particularly at optimal doping
and 2) Is Eq. (\ref{drude}) consistent with scale invariance of the
conductivity?  To answer the first question, we consult the data shown
in Fig. (\ref{optscal}) in which the optical conductivity in optimally
doped Bi$_2$Sr$_2$Ca$_{0.92}$Y$_{0.08}$Cu$_2$O$_{8+\delta}$ is plotted
assuming the Drude formula is valid.  Specifically, it is assumed that
the conductivity scales as $T^{-\mu} f(\omega/T)$ as in the Drude
formula with $1/\tau_{\rm tr}\propto T$.    Instead of a universal
function for the entire frequency range, van der Marel, et al.\cite{marel} find that $\mu=1$ for $\omega/T<1.5$
and $\mu\approx 0.5$ for $\omega/T>3$ if $f(\omega/T)$ is described by
the Drude formula.  While such a change in the exponent $\mu$ is
inconsistent with the Drude formula, there is a deeper point here.
Namely, the Drude formula is not a general consequence of scale
invariance.  The requirements that quantum criticality places on the
form of the conductivity have been explicitly worked out\cite{pchamon} under three
general assumptions, 1) the charges are critical, 2) there is only one
critical length scale and 3) charges are conserved.  Under these three
general conditions, the conductivity acquires\cite{pchamon} the form,
\beq
\label{genscal}
\sigma(\omega,T)=\frac{Q^2}{\hbar}\;T^{(d-2)/z}
\;\Sigma\left(\frac{\hbar\omega}{k_B
    T}\right),
\eeq
where $Q$ is the charge, $\Sigma$ is a universal function, and $z$ is the dynamical critical exponent defined as the exponent
that relates time and space. Technically, the dynamical exponent can be defined
as
\beq
z=\frac{d\ln\epsilon(p)}{d\ln p}
\eeq
where $\epsilon(p)$ is the quasiparticle dispersion as a function of momentum. If space and time are on equal
footing, then $z=1$.  Note absent from Eq. (\ref{genscal}) is any
additional energy scale such as the plasma frequency.  Hence, a
conductivity of the Drude form is incompatible with strict scale
invariance within a single-parameter scaling scenario. As a consequence, while the cuprates might be
describable by some kind of quantum critical scenario, it is not the
naive one in which one-parameter scaling is operative.  The root cause
of this is the mixing between the high and low energy scales which
pervades the normal state of the cuprates. 

\noindent

\subsection{Superconducting State: Color change}

\begin{figure}
\centering
\includegraphics[width=8.0cm]{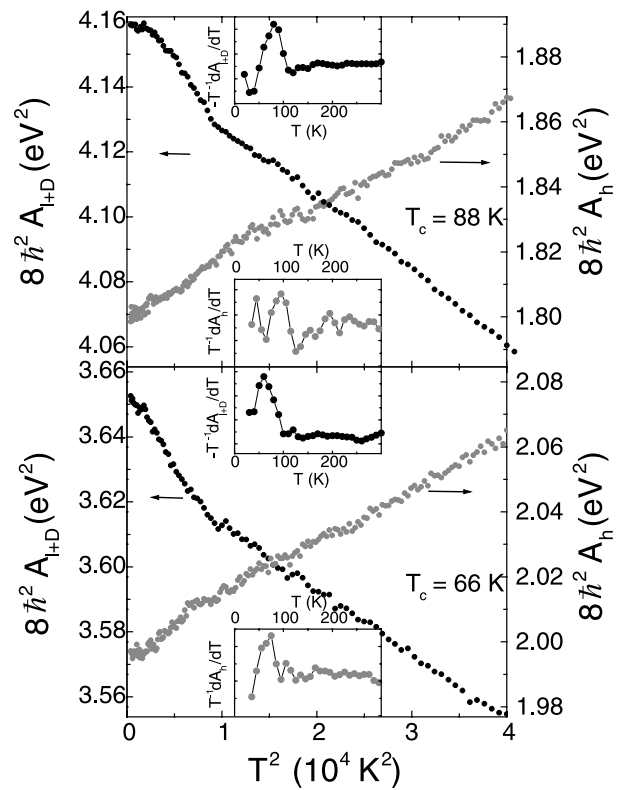}
\caption{Temperature dependence of the low-frequency spectral weight
  $A_{l+D}(T)$ and the high-frequency spectral weight $A_h(T)$ for
  optimally doped (top) and underdoped (bottom)
  Bi$_2$Sr$_2$CaCu$_2$O$_{8-\delta}$.  The insets show the derivatives
of these quantities multiplied by $T^{-1}$. Reprinted from Science,
{\bf 295}, 5563 (2002). }
\label{ahal}
\end{figure}

The phenomenon of UV-IR mixing detailed in the preceding sections
might on the surface be thought to be irrelevant once
superconductivity obtains.  That is, only phenomena on the energy
scale of the pairing interaction should be relevant to the ground
state of the supercoducting cuprates.  However, this is not the
case. R\"ubhausen, et al.\cite{rubhaussen} observed, that in underdoped
BSCO, changes occur in the optical conductivity up to 3eV or
$100\Delta$, where $\Delta$ is the superconducting order gap.  But
perhaps the changes in the 3eV range are just indicative of some
strong-coupling effect that has nothing to do with the condensation to
the superconducting state.  To answer this question, we focus on the
f-sum rule,
\beq
A=\pi n e^2/(2m)=\int_0^\infty \sigma(\omega)d\omega.
\eeq
In understanding the spectral changes in a high-$T_c$ superconductor,
it is helpful to separate $A$ into a low-energy component,
\beq
A_l=\int_{0^+}^\Omega \sigma(\omega)d\omega
\eeq
and a high-energy part
\beq
A_h=\int_\Omega^\infty \sigma(\omega)d\omega.
\eeq
The cutoff $\Omega$ is chosen so that $A_h$ contains strictly the
spectral weight associated with interband transitions.  Typically,
$\Omega/(2\pi c)=10,000 cm^{-1}$ is sufficient  to demarcate the
minimum of $\sigma(\omega)$ which demarcates the boundary between the intra-band and inter-band
transitions. The opening of a gap
opens in the optical conductivity accompanies the transition to the
superconducting state.  The spectral weight removed for
$\hbar\omega<\Delta$, $\Delta$ the superconducting gap, is transferred
to a $\delta-$function at zero frequency.  The weight in the $\delta$
function is captured by the Ferrell-Glover-Tinkham\cite{fgt}
 sum rule 
\beq
D=A_l^n-A_l^s+a_h^n-A_h^s.
\eeq
In BCS superconductors, there is no contribution to $D$ from $A_h$.
Typically, $D$ is recovered simply by integrating up to no more than
$10\Delta$.  However, the ellipsometry experiments\cite{rubhaussen} in which changes in
the dielectric function obtain for $100\Delta$ suggest otherwise for
the cuprates.  Indeed this is so.  For BSCO\cite{marel1,lupi5} both optimally and
underdoped, $A_h$ diminishes as the temperature decreases and a
compensating increase is observed for $A_l$ as depicted in Fig. (\ref{ahal}).  This indicates that it
is the loss of spectral weight in the high-energy sector that drives
the  superconducting state.  These data are also consistent with other
optical measurements which indicate that the full weight of the
$\delta-$function in the superconducting state is recovered by
integrating the optical conductivity out to
2eV\cite{bontemps} and numerical calculations\cite{scalapino} that the
frequency-dependent pairing interaction in the Hubbard model involves a non-retarded part that
arises entirely from the upper Hubbard band.  
 This color change from the visible to the
infrared implies that superconductivity in the cuprates is
fundamentally different from that in metals.  That is, in the
cuprates, superconductivity is not simply about low-energy physics on
a Fermi surface. The correct theory should
explain precisely how loss of spectral weight at high energies (2 eV
away from the chemical potential) leads
to a growth of the superfluid density. 

\noindent                                                                    

\section{Wilsonian Program for a Doped Mott Insulator}

The essence of the optical experiments on the normal state of the
cuprates is that the number of particle addition states per electron
per spin exceeds unity, in direct violation of the key Fermi liquid
tenet.  Within the Hubbard model, this state of affairs obtains
because of the dynamical mixing of the UHB and the LHB. That is, it is absent if double occupancy of bare
electrons is projected out. In this limit, $L=2x$ and $L/n_h=1$.  The key question that arises is: How can such mixing be
incorporated into a low-energy theory of a doped Mott insulator?  As
has been pointed out previously,
$L/n_h>1$\cite{sawatzky,charge2e,charge2e3}, or more generally that the
intensity of the LHB exceeds $1+x$, implies that the true
low-energy theory of a doped Mott insulator must contain more than
just electrons.  As the new
degrees of freedom arise from the mixing with doubly occupied sites,
it is reasonable that some sort of collective charge 2e excitation
should emerge. We show explicitly here how this physics arises.

\subsection{Traditional Method}

Of course, the traditional approach to constructing a
low-energy theory of a doped Mott insulator does not involve 
isolating the new degree of freedom responsible for dynamical spectral
weight transfer.  Rather, some form of degenerate perturbation theory
is used to bring the Hamiltonian into block diagonal form in double
occupancy space.  More precisely, one can perform a similarity\cite{eskes,sim2}
transformation to bring the Hubbard model into block-diagonal form, in
which the energies of a given block are approximately (in the limit of
$U\gg t$) some number times $U$.  Since the Hamiltonian is now block
diagonal, it makes sense to project to the lowest energy block.  The
resultant Hamiltonian is, at leading order, the simpler model
\beq\label{tjstuff}
H_{\rm tJ}=-t\sum_{\langle i,j\rangle}\xi_{i\sigma}^{\dagger}\xi_{j\sigma}-\frac{t^{2}}{U}\sum_{i}b_{i}^{(\xi)\dagger}b_{i}^{(\xi)}
\eeq
known as the $t-J$ model.  Here $b_i= \sum_{j\sigma} V_\sigma
\xi_{i\sigma}\xi_{j\bs}$ where $j$ is summed over the nearest neighbors of
$i$ and $V_\uparrow=-V_\downarrow=1$.  At half-filling, the first term
vanishes as well as the three-site hopping process from the second term.
The result is a pure spin Hamiltonian
\beq
H_{\rm Heis}=J\sum_{ij}S_i S_j
\eeq
where each spin operator is a product of fermion
bi-linears, $S=\Psi^\dagger\Psi$, with
\beq
\Psi = \left(\begin{array}{cc}
                            c_\ua & c_\da \\
                            c^\dagger_\da & -c^\dagger_\ua
                            \end{array} \right).
\eeq
Such a spin model is clearly invariant under the transformation
$\Psi\rightarrow h\Psi$ where $h$ is an SU(2) matrix with unit norm.
 By contrast, the Hubbard model lacks this local SU(2)
symmetry.  It has only a global SU(2) as a result of the hopping term
which is present even at half-filling.  While it is certainly possible
for a low-energy theory to possess different symmetries than the UV
starting point, in this case the local SU(2) symmetry is spurious as
it appears entirely
because we have ignored double occupancy.  This discrepancy has been pointed out
before\cite{affleck,fradkin} but the low-energy replacement for the
Heisenberg model that preserves the global SU(2) symmetry of the
Hubbard model was not offered. 

Strictly speaking, however, the block diagonalization procedure does
permit double occupancy of bare electrons.   This fact is hidden by
the block diagonalization procedure itself because the resultant
eigenstates, upon block diagonalization, are complicated linear
combinations of the original electronic states.  Consequently, the
operators appearing in the $t-J$ model are not the bare electrons in
the Hubbard model.  There are $t/U$ corrections that fundamentally
change the physics and account for $L/n_h>1$. As a result, a hole in
the $t-J$ model is {\it not} equivalent to a hole in the Hubbard model.
  Consequently, to extract the information
that is in the Hubbard model with the simpler $t-J$ model, the
operators must be re-transformed to the original basis which certainly
does not respect the no-double occupancy condition. This procedure is
cumbersome.  For example, if one were to substitute the transformed
electron operators in terms of the bare electrons of the Hubbard model into the
Eq. (\ref{tjstuff}), one would obtain Eq. (\ref{eq:sc1}) (see Appendix)
as the true low-energy theory of the Hubbard model.  While the first two
terms in Eq. (\ref{eq:sc1})  constitute the $t-J$ model including the
three-site hopping process, the remainder of the terms do not preserve the number of doubly
occupied sites. As
expected, the
matrix elements that connect sectors which differ by a single doubly
occupied site are reduced from the bare hopping $t$ to $t^2/U$. All
such terms arise from the fact that the transformed and bare electron
operators differ at finite $U$.  As the terms involving double
occupancy have the same amplitude as do the pure spin terms, there is
no justification for dropping them.  Hence, although $t/U$ is small,
(of order 1/10), it is sufficiently big to have important consequences
which influence experimental observables.  The payoff is that hidden in the t/U corrections are emergent low-energy
dynamics that are qualitatively different from that of the hard
projected t-J model.  In essence, there is a problem in the order of
limits\cite{choy2005}. The limits $U\rightarrow\infty$ and
 $L\rightarrow\infty$ do
not commute.  That is, the physics is sensitive to a finite length
scale for encountering a doubly occupied site.  A comparison is
helpful here between the predictions of the t-J and Hubbard models.  The only limit in which the t-J model can be solved
exactly is at the special value of the coupling $J=t$ in one spatial
dimension in which supersymmetry obtains\cite{barres}.  At this point, we can compare directly with the Hubbard
model to see if anything is lost by ignoring double occupancy.  The
most striking differences are those for the exponents governing the asymptotic fall-off
of the density correlations, $\alpha_c$, and momentum distribution
functions ($\theta$) in the $t-J$ (with $t=J$, the supersymmetric
point) and Hubbard models in $d=1$.  Here these quantities can be
obtained exactly\cite{yang2,yang,korepin} for both models using Bethe
ansatz.  In the d=1 Hubbard model, the exponent $\theta$ approaches\cite{yang,korepin}
$1/8$ asymptotically as $U\rightarrow\infty$ for any filling.  By
contrast in the $t-J$ model\cite{yang2}, $\theta$ is strongly dependent on doping
with a value of $1/8$ at half-filling and vanishing to zero as $n$
decreases.  More surprising, the exponent $\alpha_c$ remains pinned\cite{yang2,yang} at
$2$ for the $U\rightarrow\infty$ limit of the Hubbard model at any
filling.  In fact, at any value of $U$, $\alpha_c=2$\cite{yang2,yang} in the dilute
regime of the Hubbard model in $d=1$.  In the $t-J$ model\cite{yang2} ($t=J$),
$\alpha_c$ starts at $2$ at $n=1$ and approaches a value of $4$ at
$n=0$.  Note, $\alpha_c=4$ is the non-interacting value.   That is, in
$d=1$ in the dilute regime,
density correlations decay as $r^{-2}$ in the $U\rightarrow\infty$ Hubbard
model and as $r^{-4}$ in the $t-J$ (t=J).  This discrepancy is a clear indicator that relevant
low-energy physics is lost if double occupancy of bare electrons is projected out in the parameter range considered here.
Supposedly, this is captured by the $t/U$ corrections to the electron operators
in Eq. (\ref{trans}). While it might not be surprising that $t/U$
corrections are important for $U/t=2$ as in the supersymmetric t-J
model, similar discrepancies have been noted even for $U/t=10$ in a
direct comparison between the density of states of the t-J and Hubbard
models.  The conclusion\cite{leung1992} of such studies is 
that even at strong
coupling values of $U/t$, the t-J and Hubbard models yield similar
density of states for the lower Hubbard band only at dilute fillings
starting around $n=1/2$, considerably
far away from the $n=1$ point relevant to the cuprates.  Perhaps the
limitations of the $t-J$ model in the context of the cuprates are best
summarized by one of its chief practioneers, P. W. Anderson\cite{anderson}: ``From this point of
view, it is nonsense to consider J values which are larger than some
small limiting value.  All the theoretically exciting possibilities
proposed by various advocates of the t-J model appear in this
unphysical regime and relate to no real physical system. The spate of
gauge theories, phase-separation theories, etc. which have their
existence in this large-J region are therefore almost devoid of
physicality.''

\subsection{Exact Theory: Charge 2e Boson}

The remedy is to perform the Wilsoninan\cite{wilson1972} procedure
exactly for the simplest model of a doped Mott insulator that captures
dynamical spectral weight transfer, namely the Hubbard model.  The
result will be a theory that contains all the degrees of
freedom needed to capture the low-energy spectrum of a doped Mott insulator. Indeed it is dynamical spectral weight transfer that makes the
construction of a low-energy theory of a doped Mott insulator
non-trivial. In the exact theory, such effects are mediated by a
collective charge 2e boson.  The procedure is as follows. Consider
hole-doping a Mott insulator.  The high energy scale is now the upper
Hubbard band and hence must be integrated out to acquire the true
low-energy theory.  We introduce an elemental field which
represents the creation of excitations in the upper Hubbard band.   In this
regard, the slave-boson\cite{kotliar1998} procedure has been used.
The key idea here is to construct a model which has the same matrix elements as
does the Hubbard model in which the interaction term is replaced by $U
d_i^\dagger d_i$\cite{kotliar1998}, where $d_i$ is purely bosonic.
One cannot do this, however, without simultaneously introducing\cite{kotliar1998}
three other bosonic fields and three other Lagrange multipliers.
The low-energy theory is obtained by
integrating over the $d_i$ field as it has a mass $U$.  Since the
$d_i$ field really does represent the creation of double occupancy,
integrating it out also gets rid of double occupancy.  Hence, this
procedure does not retain dynamical spectral weight
transfer, the very feature we are trying to isolate by the Wilsonian
procedure.  Simply, this procedure does too much. Although auxillary
fields are present that represent strictly singly occupied physics,
there is no doube occupancy left when $d_i$ is integrated out.  We
outline a procedure that does precisely what is demanded on a
Wilsonian account but no more.

The solution is to expand the Hilbert space so that the field associated
with the UHB, and hence with mass $U$, represents the creation of double occupancy only when a
constraint is imposed. Unconstrained, it mediates whatever physics transpires in the UHB.  The true low-energy theory will then
correspond to integration over the new field without imposing the
constraint.  As a consequence, the low-energy theory will explicitly
permit double occupancy and hence have the essential ingredients to
describe dynamical spectral weight transfer.  In so doing, the theory
will contain the constraint parameter which permitted an identification of the
new field with the creation of double occupancy in the first place.  This procedure is
analogous to that used by Bohm and Pines\cite{bp1959} in their
derivation of plasmons. In their classic derivation, Bohm and 
Pines\cite{bp1959} expanded the Hilbert space to include a collection of low-energy oscillators but
a constraint was imposed such that when it was solved, the original electron
gas Hamiltonian was recovered.  When the constraint was relaxed, the plasmon emerged
as a new propagating degree of freedom.  Our construction mirrors
theirs in essence.

To this end, we extend the Hilbert space of the Hubbard model to
include on each site a new fermionic oscillator, $D_i$, which will
represent the physics of the UHB.  Through a
constraint, $D_i^\dagger$ will represent the creation of double
occupancy.   Imposing such a constraint requires a trick because double
occupancy transforms as a boson as it involves the product of two
fermionic operators.  At the same time, there can only be one double
occupancy per site. Hence, double occupancy has both fermionic and
bosonic character.  Dealing with this dual character requires a trick
from supersymmetry.  Let us recall how supersymmetry was first
introduced into string theory.  In its earliest form, string theory was a
theory entirely of bosonic modes represented by some field
$X_\mu(\alpha)$, where $\alpha$ is the distance along the string and
$\mu$ are the cartesian coordinates.  Ultimately to make this theory
applicable to anything real, such as QCD, fermions had to be
included. To solve this problem, Ramond\cite{ramond} promoted the
Clifford matrices to the level of a field and defined a super-field
\beq
X_\mu(\alpha,\theta)=X_\mu(\alpha)+\theta\gamma_\mu(\alpha)
\eeq
where $\theta$ is a non-commuting complex Grassmann parameter and $\gamma_\mu(\alpha)$
represent the Clifford matrices.  Since $\theta^2=0$, the second
term in $X_\mu(\alpha,\theta)$ transforms as a boson.  Hence, fermions
can be included in a bosonic theory simply by multiplying a fermion by a
Grassmann.  In our problem, just the opposite procedure will be
used.  We can turn a boson, namely double occupancy, into a fermion by
multiplication with a Grassmann.  The constraint for the elemental field
$D_i$ will turn out to be roughly $\delta(D_i-\theta
c_{i\uparrow}c_{i\downarrow})$.  However, our theory is {\bf not}
 supersymmetric as $D_i$ cannot properly
be regarded as a superfield because it really does not mix bosons and
fermions as does $X_\mu(\alpha,\theta)$. The introduction of the
supersymmetric coordinate $\theta$ is simply a trick to integrate out
the high-energy physics.  Nowhere in the UV or IR limits of the theory will $\theta$ appear. This is taken care of by writing
the Lagrangian explicitly as an integration over Grassmans.  To construct the Lagrangian in this extended space, one
has to include the standard dynamical terms for the elemental fields,
$c_{i\sigma}$ and $D_i$, as well as the hopping terms such that when
the constraint is solved, one recovers the Hubbard model.  The
Lagrangian that fits the bill is 
\beq\label{LE}
{\cal L}&&=\int d^2\theta\left[\bar{\theta}\theta\sum_{i,\sigma}(1- n_{i,-\sigma}) c^\dagger_{i,\sigma}\dot c_{i,\sigma} +\sum_i D_i^\dagger\dot D_i\right.\nonumber\\
&&+U\sum_j D^\dagger_jD_j- t\sum_{i,j,\sigma}g_{ij}\left[ C_{ij,\sigma}a^\dagger_{i,\sigma}c_{j,\sigma}
+D_i^\dagger a^\dagger_{j,\sigma}c_{i,\sigma}D_j\right.\nonumber\\
&&+\left.\left.(D_j^\dagger \theta c_{i,\sigma}V_\sigma c_{j,-\sigma}+h.c.)\right]+H_{\rm con}\right]
\eeq
where
$C_{ij,\sigma}\equiv\bar\theta\theta\alpha_{ij,\sigma}\equiv\bar\theta\theta(1-n_{i,-\sigma})(1-n_{j,-\sigma})$
and $d^2\theta$ represents a complex Grassmann integration.  In this
Lagrangian, the first two terms represent the dynamics of electrons in
the lower Hubbard band and the heavy field, respectively, the third
electron hopping in the LHB, the fourth decay of the heavy field into
an electron singlet state on neighbouring sites and the fourth an
the creation of heavy field particle-hole pair accompanied by the
creation of an electron-hole pair on neighbouring sites.  The constraint Hamiltonian $H_{\rm con}$ is taken to be
\beq\label{con}
H_{\rm con} = s\bar{\theta}\sum_j\varphi_j^\dagger (D_j-\theta c_{j,\uparrow}c_{j,\downarrow})+h.c.
\eeq
and arises simply by exponentiating a $\delta-$function.  The constant $s$ has been inserted to
carry the units of energy. The precise value of $s$ will be determined
by comparing the low-energy transformed electron with that in
Eq. (\ref{trans}).  This Lagrangian was constructed so that if we solve
the constraint, that is, integrate over $\varphi$ and then $D_i$, we
obtain exactly $\int d^2\theta \bar\theta\theta L_{\rm Hubb}=L_{\rm Hubb}$, the
Lagrangian of the Hubbard model.  Hence, up to a factor of unity, our
starting Lagrangian is equivalent to the Hubbard model.

The advantage of our Lagrangian is that it permits a clean
identification of the $U$-scale physics without equating it with
double occupancy. As the theory is Gaussian in the massive field, it
can be integrated out exactly.  To accomplish this,
we define the matrix
\beq\label{eom}
 {\cal M}_{ij}=
\left(\delta_{ij}-\frac{t}{(\omega+U)}g_{ij}\sum_\sigma c_{j,\sigma}^\dagger c_{i,\sigma}\right)
\eeq
and $b_{i}=\sum_{j}b_{ij}=\sum_{j\sigma} g_{ij}c_{j,\sigma}V_\sigma
c_{i,\bar\sigma}$.  We now complete the square and integrate over
$D_i$ exactly in the partition function. At zero frequency the
low-energy Hamiltonian\cite{charge2e2} is
\beq\label{HIR1}
H^{IR}_h = -t\sum_{i,j,\sigma}g_{ij}
\alpha_{ij\sigma}c^\dagger_{i,\sigma}c_{j,\sigma} +H_{\rm int}-\frac{1}{\beta}Tr\ln{\cal M}
\nonumber
\eeq
where
\beq\label{HIR}
H_{\rm int}=-\frac{t^2}U \sum_{j,k} b^\dagger_{j}
({\cal M}^{-1})_{jk} b_{k}-\frac{s^2}U\sum_{i,j}\varphi_i^\dagger
 ({\cal M}^{-1})_{ij} \varphi_j\nonumber\\
-s\sum_j\varphi_j^\dagger c_{j,\uparrow}c_{j,\downarrow}
+\frac{st}U \sum_{i,j}\varphi^\dagger_i ({\cal M}^{-1})_{ij}
b_{j}+h.c.\;\;,
\eeq
which constitutes the true (IR) limit as the high-energy scale has
been removed. This Hamiltonian appears complicated but can be handled
simply as will be seen.  In essence, the first term in $H_{\rm int}$,
which contains the spin-spin interaction and three-site hopping terms,
is irrelevant relative to the terms linear in $b$ so far as the charge
excitations are concerned.  It is this reduction that makes it
possible to isolate the propagating degrees of freedom in a doped Mott insulator.

To fix the
energy scale $s$, we determine how the electron operator transforms in
the exact theory. As is standard, we add a source term to the
starting Lagrangian which generates the canonical electron operator
when the constraint is solved. For hole-doping, the appropriate
transformation that yields the canonical electron operator in the UV
is
\beq
{\cal L}\rightarrow {\cal L}+\sum_{i,\sigma} J_{i,\sigma}\left[\bar\theta\theta(1-n_{i,-\sigma} ) c_{i,\sigma}^\dagger + V_\sigma D_i^\dagger \theta c_{i,-\sigma}\right] +
h.c.\nonumber
\eeq
However, in the IR in which we only integrate over the heavy degree of
freedom, $D_i$, the electron creation operator becomes
\beq\label{cop}
c^\dagger_{i,\sigma}&\rightarrow&(1-n_{i,-\sigma})c_{i,\sigma}^\dagger
+ V_\sigma \frac{t}{U} b_i c_{i,-\sigma}\nonumber\\
&+& V_\sigma \frac{s}{U}\varphi_i^\dagger c_{i,-\sigma}
\eeq
to linear order in $t/U$. This equation bares close resemblance to the
transformed electron operator in Eq. (\ref{trans}), as it should. In
fact, the first two terms are identical. The last term in
Eq. (\ref{trans}) is associated with double occupation. In
Eq. (\ref{cop}), this role is played by $\varphi_i$. Demanding that Eqs. (\ref{trans}) and (\ref{cop}) agree requires that $s= t$, thereby eliminating
any ambiguity associated with the constraint
field. Consequently, the complicated interactions appearing in
Eq. (\ref{eq:sc1}) as a result of the inequivalence between
the tranformed and bare fermions are replaced by a single charge $2e$ bosonic field
$\varphi_i$ which generates dynamical spectral weight transfer across the
Mott gap. While it is traditional in solid state systems in which both
bosons and fermions appear to integrate out the bosons, that would be
incorrect here.  Both the bosons and fermions are light degrees
of freedom that mediate low-energy physics.

While electron number conservation is broken in the IR, we find by
inspection of Eq. (\ref{HIR}) that a conserved low-energy charge does
exist, given by
\beq\label{Q}
Q=\sum_{i\sigma}c_{i\sigma}^\dagger c_{i\sigma}+2\sum_i\varphi^\dagger_i\varphi_i.
\eeq
Physically, $Q$ should equal the total electron filling, implying 
immediately implying that the weight of the low-energy fermionic part must be
less than the conserved charge. In fact, Eq. (\ref{Q}) gives a prescription\cite{dswtfinal} for $\alpha$,
namely the bosonic charge, if we interpret $Q$ as $1-x$ and the
fermionic quasiparticle density as $1-x'$, thereby corroborating the quasiparticle picture in Fig. (\ref{spec}b).

\subsubsection{Half-Filling: Bound doublon/holon pairs}

\begin{figure}
\centering
\includegraphics[width=8.0cm,angle=0]{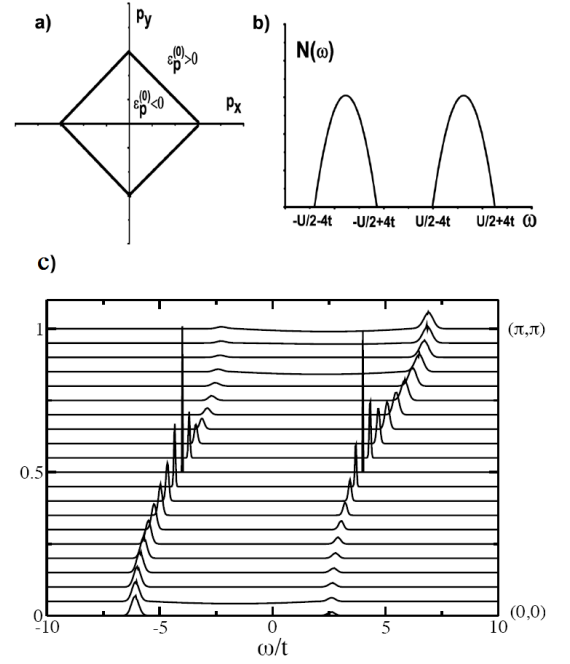}
\caption{a) Diamond-shaped surface in momentum space where the
  particle dispersion changes sign.  b) Turn-on of the spectral weight in the upper and lower Hubbard
  bands for the composite excitations as a function of energy and momentum.   In the UHB, the
  spectral density is determined to $\gamma_{\vec p}$  while for the LHB it is
  governed by $\tilde\gamma_{\vec p}$.  The corresponding operators
which describe the turn-on of the spectral weight are the composite
excitations $\varphi^\dagger cc$ (UHB) and $\tilde\varphi cc$
(LHB). The electron spectral density is determined by an overlap (see
Eq. (\ref{ov})) with
these propagating collective modes. c) Spectral function for the
electrons at $U=8t$.  Clearly shown is the gap in the spectrum and
non-zero spectral weight at all momenta in the first Brillouin zone.}
\label{mottgap}
\end{figure}

As it is the dynamical part of the spectral weight that is
intimately connected with the pseudogap, it stands to reason that the
ultimate role of the charge 2e boson is to mediate bound charge
excitations. Physically, it is reasonable that the charge 2e boson
can only influence the dynamics at low energies through the formation
of bound states because it does not have a Fock space of its own.
That is, once the heavy field is integrated out, the
Hilbert space is that of the Hubbard model.  Immediate evidence
that this state of affairs, namley bound-state formation, obtains arises from the equivalent theory
at half-filling.  At half-filling, both the upper and lower Hubbard bands
reside at high energy and hence must be integrated out.  This can be
accomplished by introducing an additional fermionic field that
represents the creation of excitations in the LHB.  Its associated
Lagrange multiplier will be $\tilde\varphi_i$.  Unlike the theory at
finite doping, the theory\cite{charge2e1,charge2e3,charge2e4} at half-filling will not have the pesky
fermionic matrix, ${\cal M}$ and as a result, no bare fields will have dynamics. The exact low-energy
Lagrangian\cite{charge2e1,charge2e3} that arises from this procedure,
\beq\label{huv}
L^{\rm hf}_{\rm IR}&=& 2\frac{|s|^2}{U}|\varphi_\omega|^2+2\frac{|\tilde s|^2}{U}|\tilde\varphi_{-\omega}|^2+ \frac{t^2}{U} |b_\omega|^2\label{eq:actlineone}\\
&&+s\gamma_{\vec p}^{(\vec k)}(\omega)\varphi^\dagger_{\omega,\vec
  k}c_{\vec k/2+\vec p,\omega/2+\omega',\uparrow}c_{\vec k/2-\vec
  p,\omega/2-\omega',\downarrow}\nonumber\\
&+&\tilde s^*\tilde\gamma_{\vec p}^{(\vec k)}(\omega)\tilde\varphi_{-\omega,\vec k}c_{\vec k/2+\vec p,\omega/2+\omega',\uparrow}c_{\vec k/2-\vec p,\omega/2-\omega',\downarrow}\nonumber\\&+&h.c.,\label{eq:actlinetwo}
\eeq
contains two bosonic fields with charge $2e$ ($\varphi^\dagger$) and $-2e$ 
($\tilde\varphi$). These bosonic modes are collective degrees of
freedom, not made out of the elemental excitations.  They represent dynamical mixing with $U-$scale physics,
namely, the contribution of double holes ($-2e$) and
double occupancy ($2e$) to any state of the system.   Here
$s$ and $\tilde s$ are constants with units of energy, all
operators in Eq. (\ref{huv}) have the same site index, repeated indices
are summed both over the site index and frequency, $\omega$,
$c^\dagger_{i\sigma}$ creates a fermion on site $i$ with spin
$\sigma$, 
\beq\label{fb}
b_{\vec k}=\sum_{\vec p}\varepsilon^{(\vec k)}_{\vec p}\ c_{\vec k/2+\vec p,\uparrow}c_{\vec k/2-\vec p,\downarrow},
\eeq
and the dispersion is given by $\varepsilon^{(\vec k)}_{\vec p}=4\sum_\mu\cos(k_\mu a/2)\cos(p_\mu
a)$, where $\vec k$ and $\vec p$ are the center of mass and relative
momenta of the fermion pair.   The coefficients 
\beq
\gamma_{\vec p}^{(\vec k)}(\omega)&=&\frac{-U+t\varepsilon_{\vec p}^{(\vec k)}+2\omega}{U}\sqrt{1+2\omega/U}\nonumber\\
\tilde\gamma_{\vec p}^{(\vec k)}(\omega)&=&\frac{U+t\varepsilon_{\vec p}^{(\vec k)}+2\omega}{U}\sqrt{1-2\omega/U}
\eeq
play a special role in this theory as they account for the turn-on of
the spectral weight.  At the level of a Lagrangian, the vanishing of
the coefficient of a quadratic term defines the dispersion of the
associated particle.  All the terms which are naively quadratic,
Eq. (\ref{eq:actlineone}), possess constant coefficients and hence we
reach the conclusion that there are no bare propagating bosons or
electrons.  In fact, it is the vanishing of the pesky ${cal M}$ matrix
in the half-filled system that leads to this state of affairs.  This
implies that there is no spectral weight of any kind. 
 However, a Mott insulator has spectral weight as depicted in Fig. (\ref{fig2}).  Consider the second line of the Lagrangian, Eq. (\ref{eq:actlinetwo}).  Appearing here are two
interaction terms, which describe composite excitations, whose coefficients can vanish.  These operators
might then  be thought of as the kinetic terms for composite excitations mediated
by the charge $\pm 2e$ bosonic fields (loosely speaking, we might
think of this as occurring because of the formation of bound states).
Such an interpretation is warranted
because the spin-spin interaction and all higher-order operators contained in the $|b|^2$ term
are at least proportional to $a^4$ and hence are
all sub-dominant to the composite interaction terms.  Here $a$ is the
lattice constant.  Consequently,
at the level of the Lagrangian, the turn-on of the spectral weight is
governed by the vanishing of the coefficients of the coupled
boson-fermion terms.  Fig. (\ref{mottgap})
shows explicitly that the vanishing of $\gamma$ and $\tilde\gamma$
leads to spectral weight which is strongly peaked at two distinct
energies, $\pm U/2$.  Each state in momentum space has spectral weight
at these two energies.  The width of the bands is $8t$. The particles
which give rise to the turn-on of the spectral weight are composite
excitations or the bound states of the bosonic and fermionic degrees
of freedom determined by the interaction terms $\varphi^\dagger cc$
and $\tilde\varphi cc$.   In the terms of the variables appearing in
the Hubbard model, the composite excitations correspond to bound
states of double occupancy and holes as has been postulated
previously\cite{castellani} to be the ultimate source of the gap in a
Mott insulator.  Interestingly, the slaved-particle approach\cite{castellani2}
 on the large $N$ limit of the infinite-$U$ Hubbard model also
finds that the  Mott gap originates from a gap in the spectrum of an
auxillary boson.  In so far as they generate the spectral
weight, the interaction terms can be thought of as the kinetic terms
in the low-energy action.  The gap (Mott gap) in the spectrum for the
composite excitations obtains for $U>8t$ as each band is centered at
$\pm U/2$ with a width of $8t$. Fig. (\ref{mottgap}) demonstrates that
the transition to the Mott insulating state
found here proceeds by
a discontinuous vanishing of the spectral weight at the chemical
potential to zero but a continuous evolution of the Mott gap as is
seen in numerical calculations\cite{imada} in finite-dimensional lattices but not
in the $d=\infty$\cite{dinfty} solution. 

In terms of the bare electrons, the overlap with the composite
excitations determines the Mott gap.  To determine the overlap, it is tempting to complete the
square on the $\varphi^\dagger cc$ term bringing it into a quadratic form,
$\Psi^\dagger\Psi$ with $\Psi=A\varphi+B cc$.  This would lead to
composite excitations having charge $2e$, a vanishing of the overlap
and hence no electron spectral
density of any kind.  However, the actual excitations that
underlie the operator $\varphi^\dagger cc$ correspond to a
linear combination of charge $e$ objects, $c^\dagger$ and $\varphi^\dagger c$.  In terms of the UV variables, the latter can be
thought of as a doubly occupied site bound to a hole.  To support this
claim, we re-do the procedure quoted above for generating the electron
operator.  At half-filling\cite{charge2e1,charge2e3}, the exact representation of the electron creation
\beq\label{cop1}
c_{i,\sigma}^\dagger\rightarrow \tilde c_{i,\sigma}^\dagger&\equiv& -
V_\sigma\frac{t}{U}\left(c_{i,-\sigma}b_i^\dagger + b_i^\dagger c_{i,-\sigma}\right)\nonumber\\
&+&V_\sigma\frac{2}{U}\left(s \varphi_i^\dagger + \tilde s \tilde\varphi_i\right) c_{i,-\sigma}
\eeq
is indeed a sum of two composite excitations, the first having to do
with spin fluctuations ($b^\dagger c$) and the other with
high-energy physics, $\varphi^\dagger c$ and $\tilde \varphi c$, that
is, excitations in the UHB and LHB, respectively. 
We can think of the overlap 
\beq\label{ov}
O=|\langle c^\dagger|\tilde c^\dagger\rangle\langle \tilde
c^\dagger|\Psi^\dagger\rangle|^2 P_\Psi
\eeq
in terms of the physical process of
passing an electron through a Mott insulator.  The overlap will
involve that
between the bare electron with the low-energy excitations of Eq. (\ref{cop1}), $\langle c|\tilde
c\rangle$, and the overlap with the propagating degrees of freedom,
$\langle \tilde c|\Psi\rangle$ with $P_\Psi$, the 
propagator for the composite excitations.  Because of the
dependence on the bosonic fields in Eq. (\ref{cop1}), $O$ retains
destructive interference between states above and below the chemical
potential.  Such destructive
interference between excitations across the chemical potential leads to a
vanishing of the
spectral weight at low energies\cite{sawatzky}.  Consequently, the
turn-on of the {\it electron} spectral weight cannot be viewed simply as a sum of the spectral weight
for the composite excitations.  As a result of the destructive interference, the
gap in the electron spectrum will always exceed that for the composite
excitations.  Hence,
establishing (Fig. (\ref{mottgap})) that the composite excitations display a gap is a
sufficient condition for the existence of a charge gap in the electron
spectrum.  A simple calculation, Fig. (\ref{mottgap}), of the electron spectral function at
$U=8t$ confirms this basic principle that a gap in the propagating
degrees of freedom guarantees that the electron spectrum is gapped.
Further, Fig. (\ref{mottgap}) confirms that the electron spectral
function involves interference across the Mott scale.  Consequently,
although the composite excitations are sharp, corresponding to
poles in a propagator as in Eq. (\ref{eq:actlinetwo}), the electrons are
not.  In fact, as Eq. (\ref{cop1}) lays plain, an electron is in a
linear superposition of excitations both in the lower and upper
Hubbard bands.  Consequently, in terms of the original electron
degrees of freedom, the transition to the Mott gap will involve
spectral weight transfer at energies on the Mott scale as illustrated in
Fig. (\ref{vo2}).  Alternatively, if the experimental probe were the composite or
bound states, spectral weight transfer would be absent because the
composite excitations represent the orthogonal propagating modes of a
half-filled band.

The composite excitations found here described by the vanishing of
$\gamma^{\vec k}_{\vec p}$
and $\tilde\gamma^{\vec k}_{\vec p}$ are the propagating degrees of
freedom in a Mott insulator.  They are orthogonal in the sense that
they never lead to a turn-on of the spectral weight for the composite
excitations in the same energy
range.  This analysis demonstrates that the spin-spin interaction,
contained in the $|b|^2$ term, plays a spectator role in the generation of the Mott
gap. Nonetheless, there is
a natural candidate for the antiferromagnetic order, namely
$B_{ij}=\langle g_{ij}\varphi^\dagger_i
c_{i,\uparrow}c_{j,\downarrow}\rangle$. The vacuum
expectation value of this quantity is clearly non-zero as it is easily
obtained from a functional derivative of the partition function with
respect to $\gamma_{\vec p}$. Such an antiferromagnet has no
continuity with that of weak-coupling theory.  Hence, both the Mott
gap and subsequent antiferromagnetic order emerge from composite
excitations that have no counterpart in the original UV Lagrangian but
only become apparent in a proper low-energy theory in which the
high-energy degrees of freedom are explicitly integrated out. In the
doped state, a similar state of affairs obtains.

\subsubsection{Experimental Consequences}

As Eq. (\ref{cop}) lays plain, the fermionic operator at low energies
is a linear superposition of two excitations.  The first is simply the
standard excitation in the LHB, $(1-n_{i,-\sigma})c_{i,\sigma}^\dagger$ ($n_{i,-\sigma}c_{i\sigma}$ in
the UHB for electron doping) with a renormalization from spin
fluctuations (second term).  The second is a new charge $e$ excitation,
$c_{i,-\sigma}{\cal M}_{ij}^{-1}\varphi_j^\dagger$. In the lowest
order in $t/U$, our theory predicts that the new excitation
corresponds to $c_{i,-\sigma}\varphi_i^\dagger$, that is, a hole bound
to the charge $2e$ boson.  This extra charge e state mediates
dynamical (hopping-dependent) spectral weight transfer across the Mott
gap. A saddle-point analysis will select a particular solution in
which $\varphi_i$ is non-zero.  This will not be consistent with the
general structure of Eq. (\ref{cop}) in which part of the electronic
states are not fixed by $\varphi_i$.  Similarly, mean field theory in
which $\varphi_i$ is assumed to condense, thereby thwarting the
possibility that new excitations form, is also inadequate.  The
procedure outlined in the second Appendix circumvents these problems
and preserves the integrity of Eq. (\ref{cop}). 
 
\begin{figure}
\centering
\includegraphics[width=6.cm,angle=270]{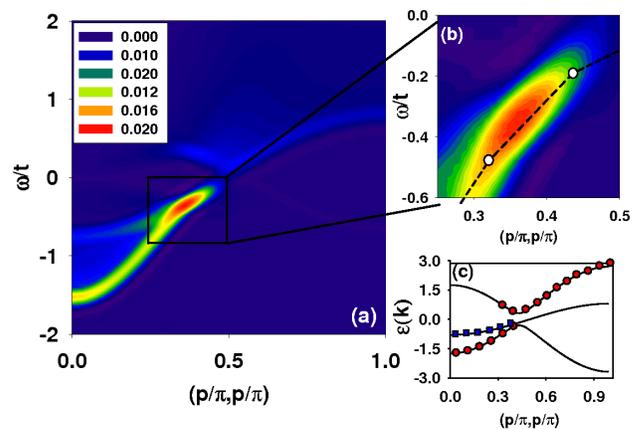}
\caption{(a) Spectral function for filling $n=0.9$ along the nodal direction.  The intensity is indicated by the color scheme.  (b) Location of the low and high energy kinks as indicated by the change in the slope of the electron dispersion.  (c) The energy bands that give rise to the bifurcation of the electron dispersion.}\label{specf1}
\end{figure}

\begin{figure}
\centering
\includegraphics[width=8.cm]{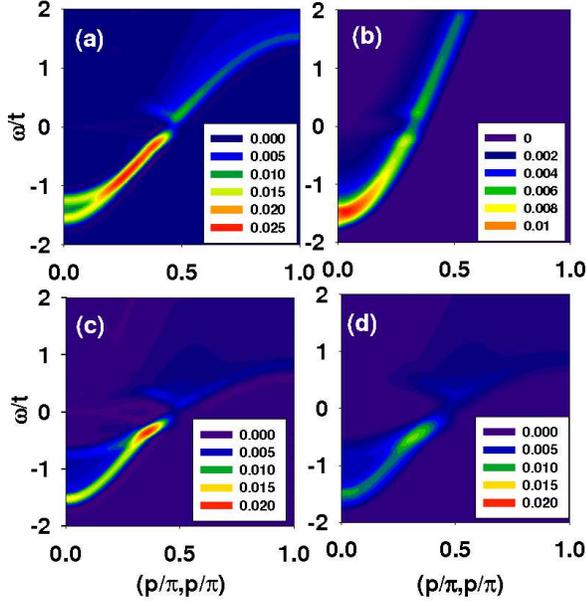}
\caption{Spectral function for two different fillings (a) $n=0.8$ and (b) $n=0.4$ along the nodal direction. The absence of a splitting in the electron dispersion at $n=0.4$ indicates the bifurcation ceases beyond a critical doping. The spectral functions for two different values of the on-site repulsion,
  (c)$U=10t$ and (d)$U=20t$ for $n=0.9$ reveals that the high-energy kink and the splitting of the electron dispersion have at best a weak dependence on $U$.  This indicates that this physics is set by the energy scale $t$ rather than $U$.}\label{specf2}
\end{figure}

The resultant spectral function $U=10t$  is
displayed in Figs. (\ref{specf1}) and (\ref{specf2}).  First, a
low-energy kink is present in the electron dispersion for a wide range
of doping.  In fact, more than one kink exists as is evident from the
enlarged region, Fig. (\ref{specf1}a): 1) one at roughly $0.2t\approx
100meV$ and the other at $0.5t\approx 250 meV$. To pin-point the
origin of these kinks, we treated the mass term of the boson as a
variable parameter and verified that the low-energy kink is determined
by the bare mass.  In the low-energy theory, the bare mass of the
boson is $t^2/U$, independent of doping. Both the doping dependence
and energy of this kink are consistent with experiment\cite{lenkink}.
While phonons\cite{lenkink} and spin-fluctuations\cite{spfkink1,spfkink2}
have been invoked to explain the low-energy kink, the hidden charge 2e
boson offers a natural explanation within the strong correlation
physics of the Mott state. 

At sufficiently high doping (see Figs. (\ref{specf2}a) and
(\ref{specf2}b)), the high-energy kink disappears. Experimentally the
origin of the high-energy kink is hotly debated.  In fact, some\cite{doubt} doubt
its intrinsic importance, attributing it to an extraneous matrix
element effect.  In the initial experiments by Graf, et
al.\cite{graf2007}, the high-energy kink is accompanied by a splitting
of the electron dispersion into two branches\cite{graf2007}. The two
branches were interpreted as evidence for spin-charge
separation. As evident from Fig. (\ref{specf1}), the high-energy kink is
proceeded by a bifurcation of the electron dispersion below the
chemical potential into two branches. 

The energy difference between  the two branches achieves a maximum at
$(0,0)$ as is seen experimentally.   A computation of the spectral
function at $U=20t$ and $n=0.9$ reveals that the dispersion as well
the bifurcation still persist.  Further, the magnitude of the
splitting does not change, indicating that the energy scale for the
bifurcation and the maximum energy splitting are set by $t$ and not
$U$. The origin of the two branches is captured in
Fig. (\ref{specf1}c).  The two branches below the chemical potential
correspond to the standard band in the LHB (open square in
Fig. (\ref{specf1}c) on which $\varphi$ vanishes and a branch on which
$\varphi\ne 0$ (open circles in Fig. (\ref{specf1}c).  The two
branches indicate that there are two local maxima in the integrand in
Eq. (\ref{eqG}). One of the maxima, $\varphi=0$, arises from the
extremum of $G(k,\omega,\varphi)$ whereas the other, the effective
free energy (exponent in Eq. (\ref{eqG})) is minimized ($\varphi\ne
0$). Above the chemical potential only one branch survives.  The split
electron dispersion below the chemical potential is consistent with
the composite nature of the electron operator dictated by
Eq. (\ref{cop}).  At low energies, the low-energy fermions are linear
superpositions of two states, one the standard band in the LHB
described by excitations of the form,
$c_{i\sigma}^\dagger(1-n_{i\bar\sigma})$ and the other a composite
excitation consisting of a bound hole and the charge 2e boson,
$c_{i\bar\sigma}\varphi_i^\dagger$. The former contributes to the
static part of the spectral weight transfer (2x) while the new charge
e excitation gives rise to the dynamical contribution to the spectral
weight transfer. Because the new charge e state is strongly dependent on the
hopping it should disperse as is evident from
Fig. (\ref{specf2})
and also confirmed experimentally. The operator that describes this
excitation is given by Eq. (\ref{bogoqp}) and it constitutes the
propagating degree of freedom in a doped Mott insulator as it generates a
pole in the composite particle Green function
(Eq. (\ref{gfinal})). In the composite basis, the electron are not
sharply defined.

\begin{figure}
\centering
\includegraphics[width=4.5cm, angle=270]{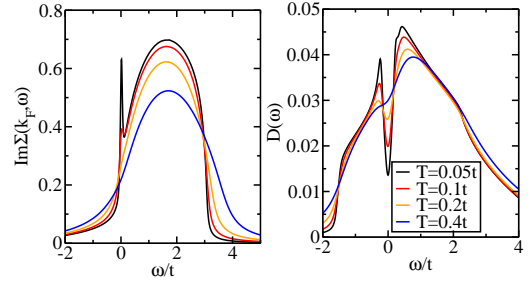}
\caption{The imaginary part of the self energy as the function of
  temperature for $n=0.7$. A peak is developed at $\omega=0$ at low
  temperature which is the signature of the opening of the
  pseudogap. The density of states explicitly showing the pseudogap is
shown in adjacent figure.}\label{self}
\end{figure}

Because the charge 2e boson is a local
non-propagating degree of freedom, the formation of the composite
excitation,
$c_{i\bar\sigma}\varphi^\dagger$ leads to a pseudogap at the chemical
potential.  The spectral functions, Fig.
 (\ref{specf2}) at $n=0.9$ and $n=0.8$ reveal an
absence of spectral weight at the chemical potential.  In the
strongly overdoped regime, spectral weight emerges at the chemical
potential and the pseudogap vanishes.   The formation of a gap
in a single-band system, requires a vanishing of single-particle Green
function.  This is mediated by a divergence of the electron
self-energy along a connected surface in momentum space.  Computed in
Fig. (\ref{self}) is the imaginary part of the self energy at different
temperatures.   At low temperature ($T\leq t^2/U$),  the imaginary part of the self-energy at the non-interacting Fermi surface develops a peak at $\omega=0$.  At $T=0$, the peak leads to a divergence.  This is consistent with the opening of a pseudogap. As we have pointed out earlier\cite{myzeros}, a pseudogap is properly identified by a zero surface (the Luttinger surface) of the single-particle Green function. This zero surface is expected to preserve the Luttinger volume if the pseudogap lacks particle-hole symmetry as shown in the second of the figures in Fig. (\ref{self}). 

\begin{figure}
\centering
\includegraphics[width=6.cm]{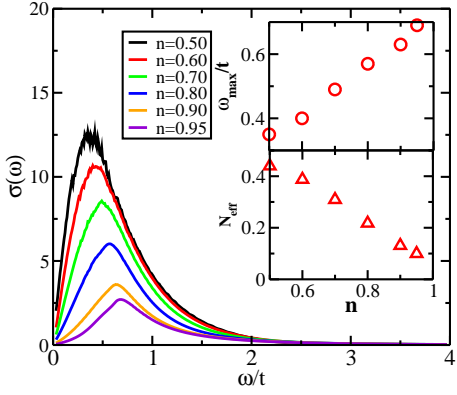}
\caption{Optical conductivity as a function of electron filling, $n$, with Drude weight at origin subtracted. The peak in the optical conductivity represents the mid-infrared band.  Its origin is mobile double occupancy in the lower-Hubbard band.  The insets show that the energy at which the MIB acquires its maximum value, $\omega_{\rm max}$ is an increasing function of electron filling.  Conversely, the integrated weight of the MIB decreases as the filling increases.  This decrease is compensated with an increased weight at high (upper-Hubbard band) energy scale. }\label{optcond1}
\end{figure}

Using this approach, we are also able to address the origin of the
mid-infrared band (MIB) in the optical conductivity. From the Kubo
formula,
\beq\label{cond}
\sigma _{xx} (\omega ) &=& 2
\pi e^2
\int d^2 k\int d\omega '(2t\sin k_x )^2 \nonumber\\
&& \left(  -\frac{f(\omega ')-f(\omega'+\omega)}{\omega} \right)
A(\omega+\omega ',k)A(\omega',k)\nonumber\\
\eeq
 we computed the optical conductivity.  Here $f(\omega)$ is the Fermi
 distribution and $A(\omega,\vec k)$ is the electron spectral
 function. At the level of theory constructed here, the
vertex corrections are all due to the interactions with the bosonic degrees of freedom.  Since the boson acquires dynamics only through electron motion and the
leading such term is $O(t^3/U^2)$, the treatment here should suffice
to provide the leading behaviour of the optical conductivity. Further,
to isolate the mid-infrared, we subtracted the Drude weight at the
origin. Apparent in the optical conductivity shown in
Fig. (\ref{optcond1}) is a peak at $\omega/t\approx .5t$. This
constitutes the mid-infrared band.  From the inset, we see that the
frequency, $\omega_{\rm max}$ at which the MIB obtains is an
increasing function of the electron function filling, whereas the
integrated weight, $N_{\rm eff}(\Omega)$, is a decreasing function as
the filling increases, both of which are in agreement with experiment. We set the integration cutoff to
$\Omega_c=2t=1/m^\ast$.  However, it does not vanish at half-filling.
Experimentally, $N_{\rm eff}$ also does not vanish when extrapolated
to half-filling. The persistence of the the MIB at half-filling
suggests that the mechanism that causes the MIB is apparent even in
the Mott state.   We determined what sets
the scale for the MIB by studying its evolution as a function of $U$.
As is clear from Figure (\ref{optcond1}), $\omega_{\rm max}$ is set essentially by the hopping matrix
element $t$ and depends only weakly on $J$. The physical processes that determine this physics are determined by the coupled boson-Fermi terms in the low-energy
theory.  The $\varphi_i^\dagger c_{i\uparrow}c_{i\downarrow}$ term has
a coupling constant of $t$ whereas the $\varphi_i^\dagger b_i$ scales
as $t^2/U$.
Together, both terms give rise to a MIB band that scales as $\omega_{\rm max}/t=0.8-2.21t/U$ (see inset of Fig. (\ref{optcond2})). Since $t/U\approx O(.1)$
for the cuprates, the first term dominates and the MIB is determined
predominantly by the hopping matrix element $t$. Within the
interpretation that $\varphi$ represents a bound state between a
doubly occupied site and a hole, second order perturbation theory with
the $\varphi_i^\dagger b_i$ term mediates the transport of hole over two sites mediated by an intervening doubly occupied site (see Fig. (1) of Ref.\cite{charge2e2}).  It is the resonance between these two states that
results in the mid-IR band.  Interestingly, this resonance persists
even at half-filling and hence the non-vanishing of $N_{\rm eff}$ at
half-filling is not evidence that the cuprates are not doped Mott
insulators as has been recently claimed\cite{millisoc}. Rather the
quantum fluctuations that are present even in the half-filled system
still persist at finite doping.  In the half-filled system, bound
states form which provide a gap to single particle excitations. In the
doped system, such excitations are only partially gapped (in momentum
space as evidenced by the V-shaped gap in Fig. (\ref{self})), giving rise
to a pseudogap. 
 
\begin{figure}
\centering
\includegraphics[width=7.5cm]{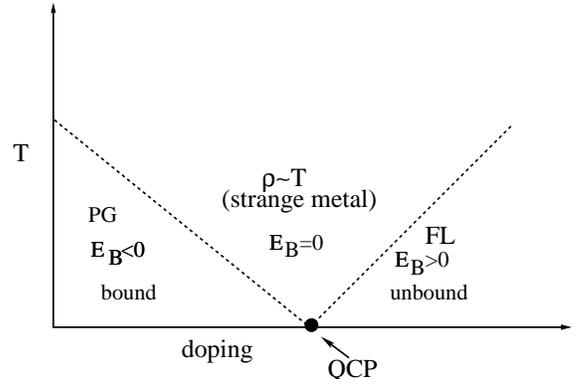}
\caption{Proposed phase diagram for the binding of the holes and bosons that result in the formation of the pseudogap phase.  Once the binding energy vanishes, the energy to excite a boson vanishes.  In the critical regime, the dominant scattering mechanism is still due to the interaction with the boson. T-linear resistivity results anytime $T>\omega_b$, where $\omega_b$ is the energy to excite a boson.  To the right of the quantum critical regime (QCP), the boson is irrelevant and scattering is dominated by electron-electron interactions indicative of a Fermi liquid. The QCP signifies the end of the binding of fermi and bosonic degrees of freedom that result in the pseudogap phase.}
\label{tlin}
\end{figure}

One of our key contentions is that the correct theory of the pseudogap
should at high temperatures explain $T-$linear
resistivity in the strange-metal regime.  Indeed this theory
has the ingredients to do this. The mechanism is simple. In the
pseudogap regime, the charge 2e boson is bound. This gives rise to an
explicit violation of the band-insulator sum rule that $L=n_h$.  
Once the $T^\ast$ line is
crossed, $\varphi$ unbinds.  Hence,  above the $T^\ast$ line,
the temperature exceeds the energy to create the charge 2e boson.
However, it still scatters off
the electrons.   The problem is now a trivial one of electron-boson
scattering above the temperature to create the boson.  This is an old
problem and the result is well known.  The resistivity scales as a
linear function of temperature. Hence, as depicted in
Fig. (\ref{tlin}), the strange metal emerges as
the unbound phase of the charge 2e boson in which critical
fluctuations are the scattering mechanism. Above the $T^*$ line,
$L=n_h$.  Hence, this mechanism makes a clear experimental prediction
that the $T^*$ line is the boundary below which $L>2x$ obtains.  A
repetition of the experiments of Chen, et al.\cite{chen} as a function
of temperature would directly falsify this claim.  Further, the
mechanism for the strange metal regime is
consistent with the implication of the scaling analysis leading to Eq. (\ref{genscal}).  Namely,
$T-$linear resistivity requires an additional energy scale absent from
a single-parameter scaling analysis.  In the exact low-energy theory, a charge 2e boson
emerges as a new degree of freedom.  While it is bound in the
pseudogap regime, its unbinding beyond a critical temperature or
doping provides the added degree of freedom to generate the anomalous
temperature dependence for the resistivity.  A further experimental
prediction of this work then is that the strange metal regime should
be populated with charge 2e excitations, without the usual diamagnetic
signal.  Shot noise measurements are ideally suited for testing this
prediction.

\section{Outlook and Predictions}

The central claim in this Colloquium is that composite excitations,
doublon-holon pairs, emerge as the
propagating degrees of freedom in the normal state of a doped Mott
insulator.  In light of other strongly
coupled problems such as QCD, this state of affairs is not surprising.
This physics arises because although double occupancy mixes into the
ground state, the empty sites which are dynamically generated are not
free to move around.  Consequently, the low-energy physics in a doped
Mott system is determined by the effective doping level $x'=x+\alpha$,
where $\alpha$ reflects the dynamical hole count.  Such physics is
mediated through the charge $2e$ boson.  At half-filling, the
band structure of the doublon-holon pairs leads to the gapped
spectrum of a Mott insulator as anticipated by Mott\cite{mott} and others\cite{castellani,numerical,kohn1}.  Because the excitations described by
$\gamma$ and $\tilde\gamma$ never share a common energy where the
spectral weight is non-zero, they can be viewed as the independent
propagating degrees of freedom in a half-filled band. 
Both the metallic state at half-filling and the strange metal are
mediated by the unbinding of the composite excitations.  The
simplest way of understanding why the charge 2e boson must be bound at
low energies, aside from the fact that it has no bare dynamics, is that once the high-energy sector is integrated out
exactly, the Hilbert space shrinks back to the Fock space of the
Hubbard model.  The charge 2e boson acting in this Hilbert space can
only mediate dynamics through binding with the elemental fields.

\begin{figure}
\centering
\includegraphics[width=9.cm]{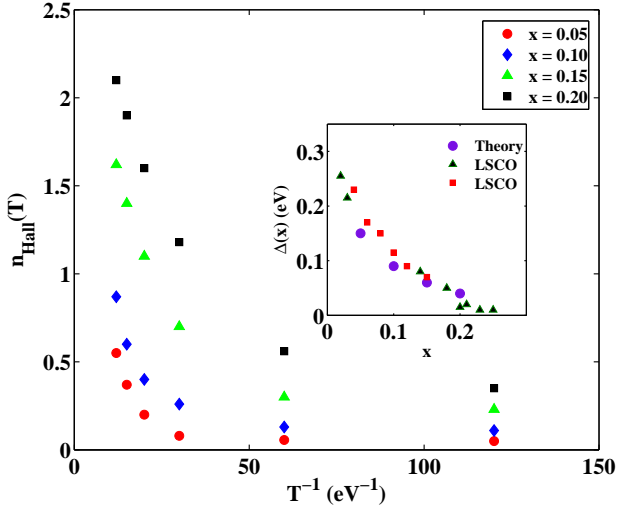}
\caption{Carrier density computed from the spectral function shown in Fig.
(\ref{specf1}).  Clearly shown are two components:  1) one independent of
temperature and scaling with the doping level and 2) the other
temperature dependent describing the new composite or bound degrees of
freedom. Fitting the carrier density to Eq. (\ref{deltac}) enables an
extraction of the gap $\Delta$ shown in the inset. The experimental values are also shown for LSCO: solid 
triangles\cite{andohall,onohall,delta1hall} and squares\cite{nishikawahall} The excellent
agreement obtained with the experimentally determined values for the
pseudogap indicates that the binding energy scale in the carrier
density is the pseudogap energy. }\label{nhallfinal}
\end{figure}

A clear indicator that the theory presented here is on the right track
is the calculation\cite{shilahall} shown in Fig. (\ref{nhallfinal}) of the carrier density as revealed by the Hall
number. Clearly shown
are two components to the carrier density, one temperature independent the other a highly temperature dependent component which
gives rise to activated behaviour in agreement with the
phenomenological fit of the experimental data in
Eq. (\ref{deltac}). In fact, the gap, $\Delta$ (see inset of
Fig. (\ref{nhallfinal})) is in excellent agreement with the
experimental data on LSCO.  Above $T^\ast$, the gapped component
vanishes.  Such two-fluid
behaviour suggests that in the effective doping level,
$\alpha$ is the
temperature-dependent component.  Any experimental probe that couples
to the low-energy excitations should be interpreted in terms of $x'$,
not the bare hole number $x$.   Several experimental predictions follow:
1) above a critical doping level, $\alpha$ should
(as a result of the unbinding of the charge $2e$ boson) and $L/n_h$
should be doping dependent, already confirmed by recent x-ray oxygen
K-edge experiments\cite{kedge}, 2) similarly, because the $T^\ast$ line
corresponds to  the unbinding of the charge $2e$ boson, angle-resolved
photoemission experiments should observe a narrowing of line shapes as
the temperautre is increased above $T^\ast$, 3) the superfluid density should 
exceed $x$ and scale as
as $x+\alpha$, already confirmed in
YBa$_2$Cu$_3$O$_{6+x}$\cite{cooper} (YBCO), 4) the inverse dielectric
function should possess two-particle hole continua\cite{charge2e2} the
second feature (starting at approximately $0.5t$) reflecting the new bound state and 5) Fermi surface volumes, that is the total volume of the hole pockets minus that of the
electron pockets,
extracted from quantum 
oscillation experiments\cite{qoscill} should be 
compared with $2(x+\alpha)$ not $2x$. The latter is particularly germane because
the Fermi surface
volumes extracted experimentally\cite{qoscill,osc1} for YBCO are not consistent with any integer
multiple of the physically doped holes.

Theoretically, several questions remain. First,
is the theory in the composite excitation picture natural in the sense
that there are no relevant perturbations to normal-state physics.  This would
require a calculation of the $\beta$-function which necessitates
rewriting the pure electron and boson interactions in terms of the composite
particles and then a subsequent renormalization group analysis.  At present
it is unclear how to proceed along these lines.  Second, can it be
quantified at what
doping level does the charge $2e$ boson decouple from the low-energy
physics and a Fermi liquid description becomes valid.  Third, what role do
the composite excitations play in the superconducting state.  Since
these excitations arise from a mixing with the high-energy sector,
they have a chance of accountng for the otherwise unexplained experimental observation that the onset of
superconductivity in the cuprates\cite{marel1} is accompanied by a depletion of
spectral weight at high energies ($U-$ scale physics) and a
compensating increase at low energies. 

\section{Acknowledgements} The structure of this Colloquium follows
the outline of a theory seminar I gave at CERN in June, 2009. I thank, T. P. Choy,
R. G. Leigh, D. Galanakis, S. Chakraborty, and T. Stanescu for the
extensive collaborations which led to the theory outlined here. I also
thank D. Basov and D. van der Marel for their insightful discussions on the
experimental sections and for their kind generosity in prepraing some
of the experimental figures.  This work was funded partially by the 
NSF DMR-0940992.

\section{Appendix: Canonical Transformation on Hubbard Model}

The goal of the perturbative approach is to bring the Hubbard model
into block diagonal form in which each block has a fixed number of `fictive' doubly occupied sites. We say `fictive' because the operators which make
double occupancy a conserved quantity are not the physical electrons but
rather a transformed (dressed) fermion we call $f_{i\sigma}$
defined below.
Following Eskes et al.\cite{eskes},
for any operator $O$, we define $\tilde O$ such that
$ O\equiv {\bf O}(c)$ and $\tilde{O}\equiv {\bf O}(f)$,
simply by replacing the Fermi operators $c_{i\sigma}$ with the
transformed fermions $f_{i\sigma}$. Note that $O$ and $\tilde O$ are only
equivalent in the $U=\infty$ limit. The procedure which makes the Hubbard model
block diagonal is now well known\cite{eskes,sim2}.
One constructs a similarity
transformation $S$ which connects sectors that differ by at most one
`fictive' doubly occupied site such that
\beq
H=e^S\tilde H e^{-S}
\eeq
becomes block diagonal, where $\tilde H$ is expressed in terms of the transformed fermions. In the new basis,
$[H,\tilde V]=0$,
implying that double occupation of the transformed fermions
is a good quantum number, and all of the eigenstates
can be indexed as such. 

Our focus is on the relationship between the physical and `fictive'
fermions. To leading order\cite{eskes} in $t/U$, the bare fermions,
\beq
c_{i\sigma}&=&e^Sf_{i\sigma}e^{-S}
\simeq f_{i\sigma}-\frac{t}{U}\sum_{\langle j, i\rangle}
\left[(\tn_{j\bs}-\tn_{i\bs})f_{j\sigma}\right.\nonumber\\
&-&\left.f^\dagger_{j\bs}f_{i\sigma}f_{i\bs}+f^\dagger_{i\bs}f_{i\sigma}f_{j\bs}\right],
\eeq
are linear combinations of
multiparticle states in the transformed basis as is expected in
degenerate perturbation theory
By inverting this relationship, we find that to leading order, the transformed operator is simply,
\beq
f_{i\sigma}\simeq c_{i\sigma}+\frac{t}{U}\sum_{j} g_{ij}X_{ij\sigma}
\eeq
where
\beq
X_{ij\sigma}=\left[(n_{j\bs}-n_{i\bs})c_{j\sigma}-c^\dagger_{j\bs}c_{i\sigma}c_{i\bs}+c^\dagger_{i\bs}c_{i\sigma}c_{j\bs}\right].
\eeq
What we would like to know is what do the transformed fermions look
like in the lowest energy sector. We accomplish this by computing
the projected operator
\beq
(1-\tn_{i\bs})f_{i\sigma}&\simeq &(1-n_{i\bs})c_{i\sigma}+\frac{t}{U}\sum_{j}g_{ij}\left[
(1-n_{i\bs})X_{ij\sigma}\right.\nonumber\\
&&\left.-X^\dagger_{ij\bs}c_{i\bs}c_{i\sigma}-c^\dagger_{i\bs}X_{ij\bs}c_{i\sigma}\right].
\eeq
Simplifying, we find that
\beq\label{trans}
(1-\tn_{i\bs})f_{i\sigma}&\simeq &(1-n_{i\bs})c_{i\sigma}+\frac{t}{U}V_\sigma
c_{i\bs}^\dagger b_i\nonumber\\
&+&\frac{t}{U}\sum_{j}g_{ij}\left[
n_{j\bs}c_{j\sigma}+n_{i\bs}(1-n_{j\bs})c_{j\sigma}\right.\nonumber\\
&&\left.+(1-n_{j\bs})\left(c_{j\sigma}^\dagger
c_{i\sigma}-c_{j\sigma}c^\dagger_{i\sigma}\right)c_{i\bs}\right].
\eeq
Here $V_\sigma=-V_{\bar\sigma}=1$ and
$b_i= \sum_{j\sigma} V_\sigma f_{i\sigma}f_{j\bs}$ where $j$ is summed over the nearest neighbors of $i$.
As is evident, the projected `fictive' fermions involve the projected
bare fermion, $(1-n_{i\bs})c_{i\sigma}$, which yields the $2x$ sum
rule plus admixture with the doubly occupied sector
mediated by the $t/U$ corrections.  These $t/U$ terms, which are
entirely local and hence cannot be treated at the mean-field level, generate the $>2x$ or
the dynamical part of the spectral weight transfer.  This physics
(which has been shown to play a significant role even at
half-filling\cite{trem}) is absent from projected models such as the
standard implementation\cite{lee,lee2} of the $t-J$ model in which
double occupancy of bare electrons is prohibited.

Consider now the low-energy Hamiltonian in the bare
electron basis. The answer in the transformed basis
is well-known\cite{eskes} and involves
the spin-exchange term as well as the three-site hopping term. Our
interest
is in what this model corresponds
to in terms of the bare electron operators which do not preserve
double occupancy. To accomplish this, we simply undo the
similarity transformation after we have projected the transformed
theory onto the lowest energy sector. Hence, the quantity of interest
is $H_{sc}=e^{-S}P_0e^SHe^{-S}P_0e^S$.  Since in the transformed
basis all such subspaces lie at least $U$ above the $m=0$
sector, it is sufficient to focus on $P_0e^SHe^{-S}P_0$. To
express $P_0e^SHe^{-S}P_0$ in the bare electron operators, we substitute
Eq. (\ref{trans}) into the first of Eqs. (14) of Eskes, et
al.\cite{eskes} to obtain
\begin{widetext}
\beq H_{sc}&=&e^{-S}P_0e^SHe^{-S}P_0e^S\nonumber\\
 &=& -t\sum_{\langle i,j\rangle}\xi_{i\sigma}^{\dagger}\xi_{j\sigma}-\frac{t^{2}}{U}\sum_{i}b_{i}^{(\xi)\dagger}b_{i}^{(\xi)}
-\frac{t^{2}}{U}\sum_{\langle i,j\rangle,\langle
i,k\rangle,\sigma}\left\{
\xi_{k\sigma}^{\dagger}\left[(1-n_{i\bar{\sigma}})\eta_{j\sigma}+\xi_{j\bar{\sigma}}^{\dagger}\xi_{i\bar{\sigma}}\eta_{i\sigma}+\xi_{i\bar{\sigma}}^{\dagger}\xi_{i\sigma}\eta_{j\bar{\sigma}}\right]+h.c.\right\}
\label{eq:sc1}
\eeq
\end{widetext}
as the low-energy theory in terms of the original electron
operators. Here, $\xi_{i\sigma}=c_{i\sigma}(1-n_{i\bar\sigma})$ and
$\eta_{i\sigma}=c_{i\sigma}n_{i\bar\sigma}$. 

\section{Appendix: Spectral Function at finite doping}

Our analysis of the half-filled
system points to a potential organizing principle of the strong
correlations.  The dynamics leading to the Mott gap are completely
independent of the spin-spin interaction contained in the term $|b|^2$.  In fact, this term is
strictly irrelevant relative to the terms of the form $\varphi^\dagger b$.
This suggests that we can drop the $|b|^2$ term, without losing
any relevant physics, as long as we are
interested in the charge dynamics in the normal state.  The resultant Lagrangian
\beq\label{lmott}
L _{\rm Mott}&=&\sum_{i,\sigma}(1- n_{i,-\sigma}) c^\dagger_{i,\sigma}\dot c_{i,\sigma}-t\sum_{i,j,\sigma}g_{ij}
\alpha_{ij\sigma}c^\dagger_{i,\sigma}c_{j,\sigma}\nonumber\\
 &-&\frac{t^2}U\sum_{i,j}\varphi_i^\dagger
  \varphi_i
-s\sum_j\varphi_j^\dagger c_{j,\uparrow}c_{j,\downarrow}
+\frac{t^2}U \sum_{i,j}\varphi^\dagger_i
b_i\nonumber\\&+&h.c.\;\;,
\eeq 
is quite simple in this case.  In obtaining
Eq. (\ref{lmott}), we retained only the leading term in the $t/U$
expansion for the fermionic matrix ${\cal M}$ and of course dropped
the $|b|^2$ term from Eq. (\ref{HIR1}). We refer to this Lagrangian as
$L_{\rm Mott}$ as it contains solely the charge dynamics. While the
predominant view\cite{lee,lee2} is that the spin-spin interaction dictates the
physics in the Mott state, simple power counting in the exact low-energy
 theory reveals otherwise.  Our analysis of the spectral function of a
 doped Mott insulator is based entirely on Eq. (\ref{lmott}).

We first assume that $\varphi$ has
no bare dynamics and is spatially homogeneous. The justification for
this assumption is simple: 1) there are no gradient terms in the
Lagrangian for the charge 2e boson and 2) its primary role is to
mix the sectors which differ in the number of doubly occupied
sites.  To proceed, we will organize the calculation of $G(k,\omega)$ by first integrating out the fermions (holding $\varphi$ fixed)
\begin{widetext}
\beq\label{geff}
G(k,\omega)= \frac{1}{Z}\int [D\varphi^*] [D\varphi] FT \left( {\int [Dc_i^*]
    [Dc_i]  c_i(t) c^*_j(0) \exp^{-\int L_{\rm Mott}[c,\varphi] dt} }\right).
\eeq
\end{widetext}

 To see what purely fermionic model
underlies the neglect of the spin-spin term in Eq. (\ref{HIR}),
we integrate over $\varphi_i$ in the partition function.  The full details of how to carry out
such an integration are detailed elsewhere\cite{charge2e1}.  The resultant Hamiltonian is not the Hubbard model but rather, 
\beq
H'=H_{Hubb}-\frac{t^2}{U}\sum_i b_i^\dagger b_i,
\eeq
a $t-J-U$ model in which the spin-exchange
interaction is not a free parameter but fixed to $J=-t^2/U$. 
That the $t-J-U$ model with $J=-t^2/U$ is equivalent to a tractable IR model, namely
Eq. (\ref{lmott}) without the spin-spin term, is an unexpected
simplification.  As Mott physics still pervades the $t-J-U$ model in the
vicinity of half-filling, our analysis should reveal the non-trivial
charge dynamics of this model. The effective Lagrangian can be diagonalized and written as
\begin{widetext}
\beq
\label{leff}
L=\sum_{k\sigma} \gamma_{k\sigma}^* \dot \gamma_{k\sigma} +  \sum_k (E_0 + E_k - \lambda_k)+\sum_{k\sigma} \lambda_k \gamma_{k\sigma}^* \gamma_{k\sigma},
\eeq
\end{widetext}
in terms of a set of Bogoliubov quasiparticles,
\beq\label{bogoqp}
\gamma_{k\uparrow}^*&=& +\cos \theta_k c_{k\uparrow}^* + \sin \theta_k c_{-k\downarrow} \\
\gamma_{k\downarrow}&=& -\sin \theta_k c_{k\uparrow}^* + \cos \theta_k c_{-k\downarrow}
\eeq
which define the propagating degrees of freedom in a hole-doped Mott insulator,
where $\cos^2\theta_k = \frac{1}{2}(1+\frac{E_k}{\lambda_k})$. Here,
$\alpha_k = 2 (\cos k_x + \cos k_y )$,
$E_0 =-(2\mu + \frac{s^2}{U})$,
$E_k =-g_t t \alpha_k -\mu$
$\lambda_k= \sqrt{E_k^2+\Delta_k^2}$
$\Delta_k= s\varphi^*(1-\frac{2t}{U}\alpha_k)$, and
$g_t=\frac{2\delta}{1+\delta}$ when $\delta=1-n \rightarrow 1-Q+2\varphi^*\varphi$ is a renormalized factor which originates from the correlated hopping
term $(1-n_{i\bar\sigma}) c_{i\sigma}^\dagger c_{j\sigma}(1-n_{j\bar\sigma})$.
Starting from Eq. (\ref{leff}), we integrate over the fermions in Eq. (\ref{geff}) to obtain,

\begin{widetext}
\beq\label{eqG}
G(k,\omega) = \frac{1}{Z}\int [D\varphi^*] [D\varphi] G(k,\omega, \varphi) \exp^{-\sum_k (E_0 + E_k - \lambda_k -\frac{2}{\beta} \ln(1+\exp^{-\beta\lambda_k}) )}
\eeq
\end{widetext}
where
\beq\label{gfinal}
G(k,\omega,\varphi)=\frac{\sin^2\theta_k[\varphi]}{\omega+\lambda_k[\varphi]} + \frac{\cos^2\theta_k[\varphi]}{\omega-\lambda_k[\varphi]}
\eeq
is the exact Green function corresponding to the Lagrangian, Eq. (\ref{leff}).
The two-pole structure of $G(k,\omega,\varphi)$ will figure prominently in the structure of the electron spectral function.
To calculate $G(k,\omega)$, we numerically evaluated the remaining
$\varphi$ integral in Eq. (\ref{eqG}). Since Eq. (\ref{eqG}) is
averaged over all values of $\varphi$, we have circumvented the
problem inherent in mean-field or saddle-point analyzes.  Physically,
Eq. (\ref{eqG}) serves to mix (through the integration over $\varphi$)
all subspaces with varying number of double occupancies into the
low-energy theory.  Hence, it should retain the full physics inherent
in the bosonic degree of freedom.

\bibliographystyle{apsrmpnurl}
\bibliography{colloqfinal}
\end{document}